\documentclass[aps,prl,twocolumn,showpacs,superscriptaddress,10pt,longbibliography]{revtex4-1}
\usepackage[T1]{fontenc}

\pdfoutput=1
\usepackage[dvipsnames,rgb,dvips]{xcolor}
\usepackage{graphicx}
\usepackage{float}
\usepackage{overpic}
\usepackage{amsmath,amssymb,amsfonts,bbm}
\usepackage{hyperref}
\makeatletter
\def\amsbb{\use@mathgroup \M@U \symAMSb}
\makeatother
\usepackage[bbgreekl]{mathbbol}
\usepackage{mathrsfs}

\newcommand{\tr}{\rm Tr\,}
\newcommand\ve[1]{\boldsymbol{#1}}
\newcommand{\ma}[1]{\ensuremath{\amsbb{#1}}}
\newcommand{\Reys}{\ensuremath{\textrm{Re}_{\rm s}}}
\newcommand{\Reyp}{\ensuremath{\textrm{Re}_{\rm p}}}
\newcommand{\St}{\ensuremath{\textrm{St}}}
\newcommand{\Sl}{\ensuremath{\textrm{Sl}}}
\newcommand{\rD}{{\rm D}}
\newcommand{\rd}{{\rm d}}

\begin{document}

\title{Angular dynamics of a small particle in turbulence}
\author{F. Candelier}
\affiliation{Aix-Marseille University - IUSTI (UMR CNRS 7343), 13 453 Marseille Cedex, France}
\author{J. Einarsson}
\affiliation{Department of Physics, Gothenburg University, SE-41296 Gothenburg, Sweden}
\author{B. Mehlig}
\affiliation{Department of Physics, Gothenburg University, SE-41296 Gothenburg, Sweden}

\begin{abstract}
We compute the angular dynamics of a neutrally buoyant nearly spherical particle immersed in an unsteady fluid. We assume that the particle is small, that its translational slip velocity is negligible, and that unsteady and convective inertia are small perturbations. We derive an approximation for the torque on the particle that determines the first inertial corrections to Jeffery's equation. These corrections arise as a consequence of 
local vortex stretching, and can be substantial in turbulence where local vortex stretching is strong and closely linked to the irreversibility of turbulence.
\end{abstract}
% \pacs{83.10.Pp,47.15.G-,47.55.Kf,47.10.-g}
% 83.10.Pp Particle dynamics (Fundamentals and theoretical)
% 47.15.G- Low-Reynolds-number (creeping) flows
% 47.55.Kf Particle-laden flows
% 47.10.-g General theory in fluid dynamics
% 05.40.-a Fluctuation phenomena, random processes, noise, and Brownian  motion
% 92.60.Mt Particles and aerosols
% 05.60.Cd Classical transport
% 45.50.Tn Collisions
% 47.27.-i Turbulent flows
% 05.40.Jc Brownian motion
% 47.55.Kf Particle-laden flows
% 47.27.eb Turbulence - Statistical theories and models
% 47.27.Gs Isotropic turbulence; homogeneous turbulence
\maketitle

The angular dynamics of small non-spherical particles in flows is often discussed in terms of Jeffery's theory \cite{jeffery1922,Par12,Gus14b,Voth16}, neglecting the effects of particle and fluid inertia. 
It is assumed that the instantaneous torque on the particle vanishes at every instant in time, so that the angular velocity of a small spherical particle equals half of the fluid vorticity at the particle position. 
But this is no longer true when the particle is so large that inertial effects become important.  
It is straightforward to take into account particle inertia 
 \cite{Marchioli2010,Gus14b,Challabotla2015,Voth15}. 
This may be a good approximation for heavy particles, but for
neutrally buoyant particles the acceleration of the surrounding
fluid cannot be neglected. How to model the effect of fluid inertia on the angular motion of a suspended particle
is an important question \cite{Voth15,Voth16}, yet difficult to answer.  Past theoretical 
studies have therefore often concentrated on simple cases, for example on the angular
dynamics of axisymmetric particles in shear flows, analysing the stability of Jeffery orbits under inertial perturbations \cite{saffman1956,subramanian2005,einarsson2015a,rosen2015d}. 

In this Letter we estimate the first inertial contributions to the angular dynamics of a nearly spherical particle  in a time-dependent and spatially varying flow. Our three most important assumptions are that the particle is  small, almost neutrally buoyant, and nearly spherical (details are given below).
We find that a spherical particle rotates at an angular velocity different from the local flow vorticity. The difference is caused by local vortex stretching. For nearly spherical particles we find additional contributions that depend in a more complex manner on the local fluid-velocity gradients. In turbulence these effects
can alter the angular particle dynamics substantially. Recently the question was raised how the dynamics of tracer particles reflects the irreversibility of turbulence 
 \cite{xu2011,jucha2014,xu2014,pumir2016}. Our results show how the {\em inertial} angular dynamics of a small particles is linked to the breaking of time-reversal 
invariance in turbulence.

{\em Formulation of the problem}.
Consider a small, nearly neutrally buoyant, spheroidal particle (with rotational symmetry axis $\ve n$) in a fluid.  
We assume that the particle is nearly spherical, its aspect ratio $\lambda$ is close to unity: $\lambda \equiv 1+\epsilon$ where $\epsilon$ is a small parameter determining the eccentricity of the particle: $\epsilon>0$ for
prolate particles, while $\epsilon<0$ for oblate particles.
We assume that the particle is small: $\kappa \equiv {2a}/{\ell} \ll 1$. Here 
$2a$ is the length of the symmetry axis of the particle, and $\ell$ is the length over which the flow can be linearised near the particle.  

We assume that inertial effects matter, but that they are weak.  Convective inertia due to the fluid-velocity gradients 
is characterised by the shear Reynolds number
$\Reys \equiv {a^2 s}/{\nu}$, where $\nu$ is the kinematic viscosity of the fluid, and $s$ measures the magnitudes of the fluid-velocity gradients, the strain rate. 
The effect of unsteady inertia depends on the time scale $\tau_{\rm c}$
that describes how fast the boundary conditions change. The ratio
of the magnitude of unsteady and convective inertia defines the Strouhal number
$ \Sl \equiv  (s \tau_{\rm c})^{-1}$. 
The effect of particle inertia on the angular dynamics is determined by the Stokes number, the ratio of the rate of change of angular momentum and the torque:
$\St  \equiv (\rho_{\rm p}a^5s)/(\tau_{\rm c}\mu s a^3) = 
({\rho_{\rm p}}/{\rho_{\rm f}}) \Reys\,\Sl$. Here $\rho_{\rm f}$ and $\rho_{\rm p}$ are the mass densities of the fluid and the particle, and $\mu = \rho_{\rm f}\nu$ is the dynamic viscosity. 

We treat the effect of inertia perturbatively, this requires $\Reys$ and $ \Reys \Sl$ to be small (but not too small, see below). 
We disregard the effect of translational slip.  This is justified for small particle Reynolds number,  $\Reyp \equiv {av_{\rm s}}/{\nu} \ll \Reys^{1/2}$. Here  $v_{\rm s}$ is the slip velocity. To ensure that it is small enough we assume that the particle
is approximately neutrally buoyant.

{\em Equations of motion}. The equations that govern the angular particle dynamics read:
\begin{equation}
\frac{\rd \ve n}{\rd   t} = 
\ve \omega \wedge  \ve n\,,\,\,\,\,\,\,\, 
\big( \ma I\cdot \tfrac{\rd \ve \omega }{\rd t }
+\tfrac{\rd \ma I }{\rd t } \cdot \ve \omega
\big) = \ve T\,.
\label{kin_momentum}
\end{equation}
Here $\ma I$ is the moment-of-inertia tensor of the particle, 
with elements 
$I_{ij} = A^I n_in_j + B^I(\delta_{ij}-n_in_j)$, $A^I$ and $B^I$ are the moments of inertia around and transverse to the axis $\ve n$, and
$\ve T$ is the hydrodynamic torque:
\begin{equation}
\label{eq:t0}
\ve T = \int_{S_p}\!\! \ve r \wedge  \bbsigma\cdot {\rm d}\ve s \,.
\end{equation}
The integral is over the particle surface $S_p$, ${\rm d}\ve s$ is the outward normal surface element,
and $\bbsigma(\ve x,t)$ is the stress tensor of the fluid at position $\ve x$, and $\ve r \equiv \ve x-\ve x_p$ where
$\ve x_p$ is the particle position.
The torque (\ref{eq:t0}) is determined by the solution of Navier-Stokes equations.  
We decompose the stress
as  $\bbsigma  = \bbsigma^\infty  + \bbsigma^{(1)}$,
where $\bbsigma^\infty$ is the stress tensor of the undisturbed Eulerian fluid velocity,
denoted by $\ve U^\infty(\ve x,t)$. The second term, $\bbsigma^{(1)}$,
is the contribution to the stress tensor from the disturbance flow.
The torque is decomposed in a similar way, $\ve T = \ve T^{\infty}+\ve T^{(1)}$.
We compute these two contributions to the torque separately.

{\em Torque due to stress of undisturbed flow.}
The undisturbed Eulerian fluid velocity $\ve U^\infty(\ve x,t)$ 
satisfies Navier-Stokes equations in the laboratory frame:
\begin{equation}
\label{eq:nsig}
\ve \nabla \cdot \bbsigma^\infty  \!= \!  \rho_{\rm f}\big[(\partial_t \ve U^\infty)_{\ve x} + (\ve U^\infty\! \cdot \!
\ve\nabla) \ve U^\infty\big] \equiv \rho_{\rm f}\tfrac{\rD}{\rD t}\ve  U^\infty\,.
 \end{equation}
The last equality defines the Lagrangian derivative along $\ve U^\infty$, 
$\ve \nabla$ denotes the spatial derivative with respect to $\ve x$, and the partial time derivative is evaluated
at fixed $\ve x$. 
The torque due to the undisturbed stress  can be expressed as a volume integral using Eq.~(\ref{eq:nsig}):
\begin{align}
\label{eq:tinfdudt}
        \ve T^\infty &= \rho_{\rm f} \int_{V_p} \rd v\,\ve r \wedge \frac{\rD \ve U^\infty}{\rD t}\,.
\end{align}
To evaluate this expression further we expand $\rD \ve U^\infty(\ve x,t)/ \rD t$  around the particle position $\ve x_{\rm p}$:
\begin{align}
\nonumber
\frac{\rD U_k^\infty}{ \rD t}\Big|_{\ve x(t)}
&=   \frac{\rD U_k^\infty}{ \rD t}\Big|_{\ve x_p(t)}
+  r_m \frac{\partial}{\partial x_m} \frac{\rD U_k^\infty}{ \rD t} \Big|_{\ve x_p}  \\
&\hspace*{-15mm}+ r_m r_n \frac{\partial}{\partial x_m}\frac{\partial}{\partial x_n}
 \frac{\rD U_k^\infty}{ \rD t} \Big|_{\ve x_p(t)} 
+ O\big(r^3/\ell^3\big)\,.
\end{align}
The components are relative to a fixed Cartesian basis in the laboratory frame.
 Substituting this expansion into Eq.~(\ref{eq:tinfdudt}) we find:
\begin{equation}
\label{eq:tinf2}
 T_i^\infty  \!=\! \rho_{\rm f}  \int_{{V}_p} \!\!\!\rd^3r \,\varepsilon_{ijk} r_j \Big(\frac{\partial}{\partial x_m}\!
 \frac{\rD U_k^\infty}{ \rD t} \Big|_{\ve x_p}\Big) r_m  + O\big({r^4}/{\ell^4}\big)\,.
\end{equation}
The order is  $O({r^4}/{\ell^4})$ because terms  odd in $\ve r$ vanish upon integration. 
The partial derivative in the integrand of Eq.~(\ref{eq:tinf2}) evaluates to
$\ve \nabla  ({\rD \ve U^\infty}/{ \rD t})|_{ \ve x_p}\!\!\!= \! ({\rD \ma A^\infty }/{ \rD  t})|_{{\ve x}_p}\!\!\! +\! {\ma A_p^\infty\cdot\ma A_p^\infty}$, 
where the elements of $\ma A_p^\infty$ are the gradients of the undisturbed fluid velocity $\ve U^\infty$  at the particle centre:
$(\ma A^\infty_p)_{ij} \equiv {\partial U^\infty_i}/{\partial x_j}|_{\ve x = \ve x_p}$.

Finally, to perform the integral in Eq.~(\ref{eq:tinf2}) we use  the definition of the moment-of-inertia tensor. We obtain:
\begin{eqnarray}
\label{eq:tinf3}
\ve T^\infty &=&    \frac{\rho_{\rm f}}{\rho_{\rm p}}\big\{\ma I\cdot \big( \tfrac{\rD \ve \Omega^\infty}{\rD t} \big|_{\ve x_p}
-  \ma S_p^\infty \cdot \ve \Omega_p^\infty\big)\\
 &&\hspace{-12mm}+   (A^I \!-\! B^I)
\big[\big(\ma S_p^\infty\cdot\ma S_p^\infty
\!+\!\ma O_p^\infty\cdot \ma O_p^\infty \!+\!   \tfrac{\rD \ma S^\infty}{\rD t}\big|_{\ve x_p}  \big) \cdot \ve n \big]\wedge \ve n\big\}\,.
\nonumber
\end{eqnarray}
Here we have decomposed the matrix $\ma A_p^\infty$ into its symmetric and antisymmetric
parts, the strain-rate matrix $\ma S_p^\infty$, and $\ma O_p^\infty$.  The matrix  $\ma O_p^\infty$
is linked to $\ve \Omega^\infty_p
\equiv \tfrac{1}{2} \ve \nabla \wedge \ve U^\infty$ through $\ma O^\infty \cdot \ve r = \ve \Omega^\infty \!\wedge \ve r$.
Eq.~(\ref{eq:tinf3}) is valid for a spheroid with
arbitrary aspect ratio
(provided that $\kappa  \ll 1$).

Using the vorticity equation $\tfrac{\rD \ve \Omega^\infty}{ \rD t} - \ma S^\infty \cdot \ve \Omega^\infty =
  \nu \nabla^2 \ve \Omega^\infty$
evaluated at the particle position, we can express the difference in the first row of Eq.~(\ref{eq:tinf3})
in terms of the Laplacian of $\ve \Omega^\infty$ which vanishes for a strictly linear flow.
In general, however, the non-linearity of $\ve U^\infty(\ve x,t)$ results
in a non-zero value of  $\nabla^2\ve \Omega^\infty$ at the particle position.
\begin{figure}
\begin{minipage}{8cm}
\begin{overpic}[width=3cm]{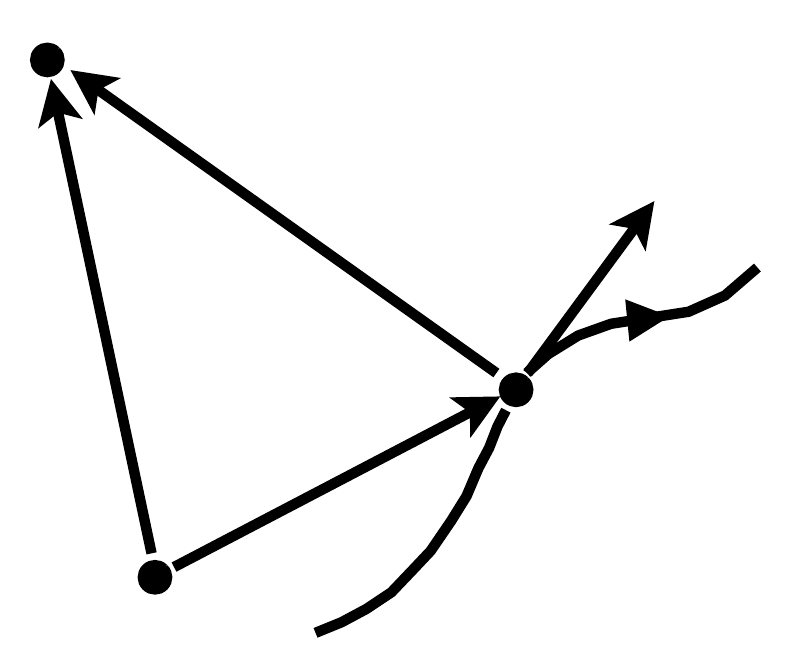}
\put(0,40){$\ve x$}
\put(8,0){$\ve O$}
\put(28,28){$\ve x_p$}
\put(40,60){$\ve r$}
\put(54,67){$\frac{\rd \ve x_p}{\rd t}\!=\!\ve U^\infty(\ve x_p,t)$}
\end{overpic}
\hspace*{12mm}\raisebox{1.15cm}{
\begin{tabular}{l}
\small Eulerian fluid \\
\small velocity   $\ve U(\ve x,t)$ \\[1mm]
\small Undisturbed Eulerian   \\
\small fluid velocity  $\ve U^\infty(\ve x,t)$\\[1mm]
\small Disturbance flow\\
$\ve w(\ve r, t)=\ve U(\ve r+\ve x_p,t)$\\
\hspace*{1.07cm}$-\ve U^\infty(\ve r+\ve x_p,t)$
\end{tabular}}
\end{minipage}
\caption{\label{fig:1} Schematic of vectors used to describe the particle motion,
$\ve x$ is a position in the laboratory frame, $\ve O$ is its origin,
and $\ve x_p(t)$ is the particle position at time $t$. The perturbation
theory leading to Eq.~(\ref{eq:result}) assumes that the particle follows the Lagrangian fluid trajectory. }
\end{figure}

{\em Disturbance torque.} We calculate the torque due to the disturbed fluid in perturbation theory, assuming that inertial effects are small
and neglecting translational slip. This dictates how the problem must be de-\-di\-men\-sio\-nali\-sed: 
$ t = \tau_c  t^\prime\,, \ve r = a \ve r^\prime\,, \ve U = sa \ve U^\prime\,,$ and $\bbsigma = \mu s \bbsigma^\prime\,.$ In the remainder of this Letter we use these dimensionless variables. To simplify the notation we drop the primes, all equations below are written in dimensionless variables.
The disturbance caused by the particle has the flow velocity
$\ve w(\ve r, t)\equiv\ve U(\ve r+\ve x_p,t)-\ve U^\infty(\ve r+\ve x_p,t)$.
It is defined to be a function of $\ve r(t)\equiv\ve x-\ve x_p(t)$ (Fig.~\ref{fig:1}). 
In dimensionless variables the disturbance problem reads:
\begin{eqnarray}
\label{eq:ns_w}
&&\nabla_{\ve r}\! \cdot \bbsigma^{(1)}\\
&&\hspace*{2mm}=\Reys\big[\Sl(\partial_t \ve w)_{\ve r}  \!+\!
	{(\ma A_p^\infty\!\!\cdot\!\ve r)}\!\cdot\!\nabla_{\ve r}\ve w \!+\!
	 \ma A_p^\infty \!\!\cdot\! \ve w \!+\!(\ve w\! \cdot\!\nabla_{\ve r}) \ve w\big]\,,\nonumber\\
&&\ve w(\ve r, t)\! =\! -(\ma A_p^\infty\cdot \ve r -\ve \omega \wedge \ve r)
\,\mbox{for}\, \ve r \in S_p\,,\nonumber\\
&&
\ve w(\ve r, t) = 0\,\mbox{ as }\,|\ve r| \to \infty\,.\nonumber
\end{eqnarray}
The partial time derivative is evaluated
at fixed $\ve r$, and  we have linearised $\ve U^\infty(\ve x,t)$
around $\ve x_p$,
$\ve U^\infty(\ve x,t) = \ve U^\infty(\ve x_p,t) +\ma A_p^\infty\cdot\ve r$.
When is it justified to use this linear form in the disturbance problem (\ref{eq:ns_w})?
The disturbance caused by the particle decays exponentially at distances
larger than the Saffman length $\ell_{\rm S} \equiv 1/\sqrt{\Reys}$,
so that we must require $\ell_{\rm S} < \ell$, where $\ell$ is the length scale over which the flow can be linearised.
This condition is more restrictive than $\kappa \ll 1 $. In other words the shear Reynolds number $\Reys$ should not be too small, because convective inertia causes
the disturbance to decay at $\ell_{\rm S}$.

We use the reciprocal theorem \cite{lovalenti1993}  to find the hydrodynamic torque on the particle, given an  \lq auxiliary{\rq} Stokes solution in 
the same geometry \cite{subramanian2005,einarsson2015a,einarsson2015b,candelier2015b}. In dimensionless variables the reciprocal theorem reads:
\begin{align}
	\ve T =\ve T^\infty
	+\ve T^{(0)}
	-\Reys\int_V \rd v \,{\ma M^{\sf T}}\cdot\ve f(\ve w)\,.\label{eq:torque_rt}
\end{align}
The first term on the r.h.s. is the torque due to the undisturbed fluid stresses, Eq.~(\ref{eq:tinf3}), expressed in dimensionless variables.
The second term is Jeffery's torque \cite{jeffery1922}: 
$\ve T^{(0)} = -\ma K\cdot (\ve \omega-\ve\Omega_p^\infty) + \ma H : \ma S_p^\infty$,
where $\ma K$ and $\ma H$ are Brenner's resistance tensors \cite{kim1991}.
The third term in Eq.~\eqref{eq:torque_rt} is the torque due to the disturbance flow beyond the Stokes approximation. The tensor $\ma {M}(\ve r)$ is determined 
by the known auxiliary Stokes solution ($\tilde{\ve u}=\ma M\cdot\tilde{\ve \omega}$), the Stokes flow around the particle rotating with angular velocity $\tilde{\ve \omega}$ in a quiescent fluid,
see supplemental material \cite{supp}.
The integration domain in (\ref{eq:torque_rt}) is the fluid volume outside the particle, and $\Reys\ve f$ is defined as the r.h.s of the first Equation~\eqref{eq:ns_w}.

To simplify the calculation of the third term in Eq.~\eqref{eq:torque_rt}  we assume
that the particle is nearly spherical \cite{candelier2015b}, 
$\lambda=1+\epsilon$, and expand in the small parameter $\epsilon$.
To order $\epsilon^2$, for example, the moments of inertia around and transverse
to the particle symmetry axis \cite{candelier2015b} read in dimensionless variables:
$A^I   \!=\! \frac{8 \pi }{15}(1\!-\!4\epsilon\!+\!6 \epsilon^2)$ and 
$B^I \!=\! \frac{8 \pi }{15}(1\!-\!3\epsilon\!+\!\frac{7}{2}\epsilon^2)$.
For small values of $\epsilon$ the contribution of the volume integral in Eq.~(\ref{eq:torque_rt})
can be evaluated. Details are given in the supplemental material \cite{supp}.

For a spherical particle ($\epsilon=0$) we find to order $O(\Reys)$:
\begin{equation}
\label{eq:torque}
\ve T =  -8\pi(\ve \omega-\ve \Omega_p^\infty)+ 8\pi\Reys \big( \tfrac{\Sl}{15}  \tfrac{\rD \ve \Omega^\infty}{\rD t} \big|_{\ve x_p}
\!\!\!\!- \tfrac{2}{5}  \ma S_p^\infty \cdot \ve \Omega_p^\infty\big)\,.
\end{equation}
We now use Eqs.~(\ref{eq:torque}) and (\ref{kin_momentum}) to determine the particle angular velocity $\ve \omega$ to order  $O(\Reys)$.
For a spherical particle $\ma I$ has the elements $I_{ij} = 8\pi\delta_{ij} /15$, in dimensionless variables.
In perturbation theory
${\rm d}\ve\omega/{\rm d}t = {\rm d}\ve \Omega_p^\infty/{\rm d}t = {\rm D}\ve\Omega^\infty/{\rm D}t|_{\ve x_p}$.
We must also require that $\rho_{\rm p}\approx \rho_{\rm f}$, so that $\Reyp$ remains small.
To order $O(\Reys)$ we find:
\begin{equation}
\label{eq:result0}
\ve \omega= \ve\Omega_p^\infty  + \tfrac{1}{15} \left(\Reys\,\Sl-\St\right) \tfrac{\rD \ve  \Omega^\infty}{\rD t}\big|_{\ve x_p}  
- \tfrac{2}{5} \Reys \ma S_p^\infty \cdot \ve \Omega_p^\infty\,.
\end{equation}
Eq.~(\ref{eq:result0}) is the main result of this Letter. We see that a very small particle in a viscous flow
rotates with half the fluid vorticity, $\ve \Omega_p^\infty$, as expected. The first inertial correction term, 
proportional to $\rD \ve  \Omega^\infty/\rD t|_{\ve x_p}$, resembles the form of the slip velocity of a small particle subject to particle inertia and to the force due to the undisturbed pressure gradients.
A small-$\St$ expansion gives 
\begin{equation}
\label{eq:vs}
\ve v_{\rm s} \!=\! \St\,(\rho_{\rm f}/\rho_{\rm p}-1)\, \rD\ve U^\infty/\rD t|_{\ve x_p}
\end{equation} for $\Sl=1$.
Maxey used this approximation to conclude that heavy particles
are centrifuged out of vortical regions in turbulence \cite{Max87,Gus16}.
The expression for $\ve v_{\rm s}$ is of the same form as the second term on the r.h.s. of Eq.~(\ref{eq:result0}), since $\Reys = (\rho_{\rm f}/\rho_{\rm p})\St$ for $\Sl=1$.  This term predicts that a particle that is slightly heavier
than the fluid rotates a little bit more slowly, because it cannot keep up with the the fluid acceleration. A lighter particle, by contrast,
rotates faster than $\ve \Omega_p^\infty$.
But these arguments disregard the $\ma S_p^\infty \cdot \ve \Omega_p^\infty$-term in Eq.~(\ref{eq:result0}).
This inertial term connects the angular particle dynamics to vortex stretching in the undisturbed flow.
This term vanishes in steady linear flows such as a simple shear \cite{lin1970,meibohm2016}, and for planar flows because vorticity is orthogonal to the flow plane.
\begin{figure}[t]
\begin{overpic}[width=8cm,clip]{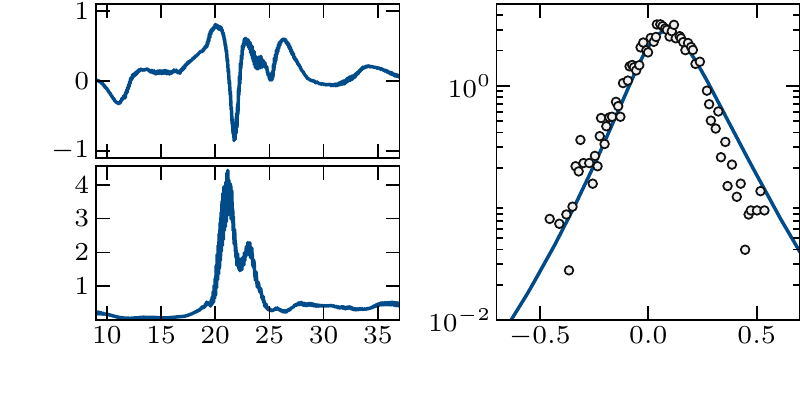}
\put(0,38){\rotatebox{90}{\colorbox{white}{$Y(t)$}}}
\put(1,10){\rotatebox{90}{$|\ve \Omega_p^\infty(t)|^2\tau_{\rm K}^2$}}
\put(30,3 ){ $t/\tau_{\rm K}$}
\put(85,3 ){ $Y$} 
\put(54,24 ){\rotatebox{90}{$P(Y)$}}
\put(14,46){\bf a}
\put(14,25.5){\bf b}
\put(64,46){\bf c}
\end{overpic}
\vspace*{-4mm}
\caption{\label{fig:2}
{\bf a} Inertial correction $Y\equiv \tau_{\rm K}{\ve \Omega_p^\infty\!\cdot\ma S_p^\infty\!\cdot\ve \Omega_p^\infty}/{|\ve \Omega_p^\infty|^2}$, Eq.~(\ref{eq:result}), along a Lagrangian path in turbulence 
(DNS at $\textrm{Re}_\lambda = 433$ using JHU turbulence database \cite{Yi2008,Yu2012}). {\bf b} Vorticity along the same path. {\bf c} Distribution of $Y$ (DNS, lines). Also shown: experimental data ($\circ$) at $\textrm{Re}_\lambda = 50$ read off 
from Fig.~2d in Ref.~\cite{guala2005}, rescaled by $\tau_{\rm K}\equiv \langle \tr \ma A^{\sf T}\cdot\ma A\rangle^{-1/2}$, see text.
The experimental value ($0.27$ s$^{-1}$) was taken from Ref.~\cite{luthi2005}.}  
\end{figure}

{\em Angular velocity and vortex stretching  in turbulence}.  Fully developed turbulent flows
exhibit large local vortex stretching rates, this is how the turbulent kinetic energy is dissipated at small scales.
Irreversibility of turbulence \cite{xu2011,jucha2014,xu2014,pumir2016} implies that the stretching rate $\ve \Omega^\infty\cdot (\ma S^\infty \cdot \ve \Omega^\infty)$ does not average to zero \cite{johnson2016,guala2005}. This matrix element is positive on average because $\ve \Omega^\infty$ tends to align with the middle eigenvector of $\ma S^\infty$  \cite{xu2011}, and its eigenvalue is positive on average. 
Eq.~(\ref{eq:result0}) shows that the inertial correction to the angular velocity of a neutrally buoyant sphere in turbulence is determined by the stretching rate,
\begin{equation}
\label{eq:result} {|\ve \omega|^2}/{|\ve \Omega^\infty_p|^2}\approx 
1 - \tfrac{4}{5}\,\Reys \, {\ve \Omega^\infty_p \cdot (\ma S_p^\infty \cdot \ve \Omega_p^\infty)}/{|\ve \Omega^\infty_p|^2}\,,
\end{equation}
and thus linked to the breaking of time-reversal invariance in 
turbulence.

But are the conditions of validity summarised above met for a neutrally buoyant particle in turbulence?
The shear rate, defined by $s = \langle \tr \ma S\cdot \ma S\rangle^{1/2}$, is of the same order as $\tau_{\rm K}^{-1}$, the inverse
of the Kolmogorov time $ \tau_{\rm K} \equiv \langle \tr \ma A^{\sf T}\cdot\ma A\rangle^{-1/2}$.  This means that the shear Reynolds number is 
of order $\Reys \sim  {a^2}/{\eta_{\rm K}^2}$,
where $\eta_{\rm K} = \sqrt{\nu \tau_{\rm K}}$ is the Kolmogorov length.
We conclude that the particle must be smaller than the Kolmogorov length
for the perturbation theory in $\Reys$ to be valid. 
For a nearly neutrally buoyant particle we can disregard gravity, and we assume that no other external forces
act on the particle. Therefore the timescale $\tau_{\rm c}$ is that of the fluid and we take 
$\tau_{\rm c} = s^{-1}\sim \tau_{\rm K}$, so that the Strouhal number is unity. The perturbation theory also requires
that the Saffman length $\ell_{\rm S} \equiv a/\sqrt{\Reys} \sim \eta_{\rm K}$ is smaller 
than the length $\ell$ over which the flow can be linearised. 
In turbulence $\ell \sim 10 \eta_{\rm K}$. So this condition
is only marginally satisfied.
Finally, translational slip is negligible when the Oseen length is much larger than the Saffman length. This is the case when $\Reyp^2\ll \Reys$. For small neutrally buoyant particles in turbulence
this condition is well satisfied.

We evaluate the inertial correction in Eq.~(\ref{eq:result}) for a neutrally buoyant particle numerically, following a Lagrangian trajectory in fully developed turbulence. We use the JHU turbulence database \cite{Yi2008,Yu2012} that contains a direct numerical simulation of forced, isotropic turbulence 
at $\textrm{Re}_\lambda = 433$. The result is shown in 
Fig.~\ref{fig:2}{\bf a}. Panel {\bf b} shows the vorticity along the same path. 
We observe the stretching of a vortex tube at $t\approx 20\,\tau_{\rm K}$. 
We see that the inertial correction to the particle angular velocity can be substantial during vortex stretching, a factor of order unity times $\Reys$. 
Panel {\bf c} shows that the distribution of the inertial correction has heavy, non-Gaussian tails that give rise to large values of the inertial correction. Comparison with experimental data  at ${\rm Re}_\lambda=50$ ($\circ$), from Fig.~2d in Ref.~\cite{guala2005}, shows that the tails are quite robust even at moderate values of $\Reys$, a consequence of the universality of dissipative-range turbulent fluctuations \cite{schumacher2014}. 

{\em Non-spherical particles}. We have computed the angular velocity also for 
nearly spherical particles ($|\epsilon|\ll 1$). Neglecting inertial effects we 
obtain an  $\epsilon$-expansion of Jeffery's equation \cite{jeffery1922}:
$\ve \omega \!=\! \ve \Omega_p^\infty + (\epsilon +\tfrac{\epsilon^2}{2})(\ve n \wedge \ma S_p^\infty\cdot \ve n) + O(\epsilon^3)$. 

The form of the first inertial corrections to Jeffery's angular velocity 
is constrained by symmetries. To linear order in $\St$ and $\Reys$
the corrections are quadratic in $\ma S_p^\infty$ and $\ve \Omega_p^\infty$,
and linear in time derivatives of $\ma S_p^\infty$ and $\ve \Omega_p^\infty$. 
The inertial corrections to $\ve \omega$ must be invariant under $\ve n \to -\ve n$, 
and $\ma S_p^\infty$ is symmetric and traceless. Only the following terms can occur in the 
$O(\epsilon\Reys)$-correction $\delta\ve \omega \equiv \epsilon\Reys\ve \omega^{(\epsilon\Reys)}
+\epsilon\St\ve \omega^{(\epsilon\St)}$ to Eq.~(\ref{eq:result0}):
\begin{eqnarray}
\label{eq:result1}\delta\ve \omega  \!&=&\!\beta_{1}  \tfrac{\rD \ve \Omega^\infty}{\rD t} \big|_{\ve x_p} 
\!\!\!\!\!+\! \beta_2  (\tfrac{\rD \ve \Omega^\infty}{\rD t} \big|_{\ve x_p}\!\!\!\!\!\!\cdot\ve n) \ve n   
+ \beta_3  \big(\tfrac{\rd \ma S_p^\infty }{\rd t}\cdot \ve n\big)\wedge \ve n  \\
&&\hspace*{-10mm}+\beta_{4} \ma S_p^\infty\cdot \ve \Omega_p^\infty \!+\! \beta_{5} [\ve n \cdot (\ma S_p^\infty\cdot \ve \Omega_p^\infty)] \ve n 
\!+\!\beta_6  [(\ve n \cdot (\ma S_p^\infty\cdot\ve n)]\ve \Omega_p^\infty\nonumber\\
&&\hspace*{-10mm}+\beta_7 (\ve n \cdot \ve \Omega_p^\infty)(\ve \Omega_p^\infty\wedge \ve n)  + \beta_8 (\ma S_p^\infty \cdot \ma S_p^\infty \cdot \ve n)\wedge \ve n 
\nonumber\\
&&\hspace*{-10mm}+ \beta_9 [\ma S_p^\infty \cdot (\ve \Omega_p^\infty\wedge \ve n)]\wedge \ve n + \beta_{10}  (\ve n \cdot \ve \Omega_p^\infty) \:\ma S_p^\infty \cdot \ve n \,.\nonumber
\end{eqnarray}
For $\Sl=1$ we find using the method described above (details in the supplemental material \cite{supp}):
\begin{align}
\beta_{1} &= -\tfrac{3\epsilon}{15}(\Reys-\St)\,,\,\,\beta_{2} = -\tfrac{\epsilon}{15}(\Reys-\St)\\
\beta_3 &= -\tfrac{\epsilon}{3} (\Reys - \tfrac{1}{5}\St) \,,\,\, \beta_{4}  = \tfrac{733\epsilon}{525}  \Reys\,,\nonumber\\
\beta_{5} &=-\tfrac{\epsilon}{15} (\tfrac{4}{35}\Reys+ \St)\,, \beta_6 = - \tfrac{\epsilon}{15}(\tfrac{30}{35} \Reys-\St)\,,\nonumber\\
\beta_7 &= -\tfrac{\epsilon}{15}(\Reys - \St)\,,\,\, \beta_8 = -\tfrac{8\epsilon}{21} \Reys\,,\nonumber\\
\beta_9 &=-\tfrac{\epsilon}{15}(4\Reys - \St)\,,\,\, \beta_{10}= \tfrac{3\epsilon}{35} \Reys\,.\nonumber
\end{align}
Eq.~(\ref{eq:result1}) shows  that the inertial corrections 
to the angular velocity of non-spherical particles depend intricately on the relative alignment of the particle symmetry axis, of the vorticity, and of the eigensystem of the strain-rate matrix \cite{Par12,Gus14b,Byron2015}. For a neutrally buoyant particle in incompressible isotropic homogeneous turbulence we can average the correction.
Since $\delta\ve \omega$ contains a factor $\epsilon$, we obtain the 
$O(\epsilon\Reys)$-result by averaging $\ve n$ uniformly over the unit sphere, $\langle n_i n_j\rangle = \delta_{ij}/3$. This gives:
$\langle \ve \Omega_p^\infty\cdot \delta\ve \omega\rangle = (4/3) \epsilon\Reys  \langle \ve\Omega_p^\infty\cdot(\ma S^\infty\cdot\ve \Omega_p^\infty)\rangle$.
On average the form of the correction is similar to that in Eq.~(\ref{eq:result0}). We see that the inertial effect is weakened for slightly prolate neutrally buoyant particles, but increased for  oblate particles.

{\em Conclusions}.
We have computed the first inertial corrections to the angular velocity of a small, approximately neutrally buoyant particle in a space- and time-dependent flow.  Our main prediction, Eq.~(\ref{eq:result}), expresses the inertial correction to the angular velocity of a small sphere in terms of a matrix element that determines vortex stretching. 
This shows that the inertial angular dynamics of a small neutrally buoyant sphere in turbulence picks up that time-reversal invariance is broken \cite{xu2011,jucha2014,xu2014,pumir2016}. 

Our results demonstrate  that convective and unsteady fluid inertia must be treated on an equal footing in turbulence. It is commonly argued that convective effects can be disregarded when the translational slip velocity $\ve v_{\rm s}$ is negligible. But we have shown here that substantial inertial corrections may arise from convective terms due to turbulent strains. Such terms are likely to be important in the translational problem too, so that the \lq Maxey-Riley\rq{} equations \cite{gatignol1983,maxey1983} cannot be used to describe the translational dynamics of small spheres in turbulence.

Particle-tracking experiments \cite{Tra15,variano2016,Voth} and particle-resolving DNS required to test the predictions in this Letter have
recently become possible \cite{Homann,Fornari}, but they are still very challenging. In order to stimulate the substantial effort required for these measurements we now give a concrete suggestion for how Eq.~(\ref{eq:result}) could be tested in either experiment or particle-resolving DNS. First, the particle should be neutrally buoyant, and its size $a$ must be smaller than the Kolmogorov length $\eta_{\rm K}$. 
Small $a$ ensures that $\Reyp^2\ll \Reys$ since $\Reyp \propto a^3$ [Eq.~(\ref{eq:vs})] while $\Reys\propto a^2$.
This may be a difficult technical requirement. It is easier to meet when the turbulence intensity is low, so that
$\eta_{\rm K}$ is larger, and Fig.~\ref{fig:2}{\bf c} shows that the effect persists for lower turbulence intensities. 
Second, to determine the vorticity of the undisturbed flow, one must measure and interpolate the flow near the particle~\cite{Tra15}. Then we suggest to consider the distribution of $|\ve \omega|^2/|\ve \Omega^\infty_p|^2-1$. The width of this distribution must approach zero for a perfect tracer particle.  Fig.~\ref{fig:2}{\bf c} shows that the first effect of inertia is to substantially widen the tails of this distribution, and to slightly shift its mean value. 
Finally, our results for nearly spherical particles indicate that disks may be more sensitive to inertial corrections than rods.

Can our results be generalised to cases where $\ve v_{\rm s}$ is not negligible? An important case is settling \cite{Gus14e,Ireland,Mathai2016,Parishani}. The settling of ice crystals, for instance, is important for rain initiation in cold turbulent clouds \cite{Pru78}.
For a spatially constant flow the effect of non-zero $\ve v_{\rm s}$ 
was analysed by Lovalenti \& Brady \cite{lovalenti1993}. Can one use their methods to compute lift forces \cite{zimmermann2011} on small particles in turbulence?  This is a difficult problem because it requires singular perturbation theory \cite{saffman1965}.
Finally, larger particles pose other problems: they sense inertial-range turbulent fluctuations \cite{Par14}, and wakes may affect their dynamics \cite{Mat15}. 

\acknowledgments{
{\em Acknowledgments}. We thank Howard Stone for discussions and
for making the preprint \cite{stone2016} available to us that considers a related question (the steady-state limit of the problem considered here).
This work was supported by Vetenskapsr\aa{}det [grant number 2013-3992], Formas [grant number 2014-585],
and by the grant \lq Bottlenecks for particle growth in turbulent aerosols\rq{} from the Knut and Alice Wallenberg Foundation, Dnr. KAW 2014.0048.
 The numerical results in Fig.~2 use data from the JHU turbulence database \cite{Yi2008,Yu2012}.}


\begin{thebibliography}{48}%
\makeatletter
\providecommand \@ifxundefined [1]{%
 \@ifx{#1\undefined}
}%
\providecommand \@ifnum [1]{%
 \ifnum #1\expandafter \@firstoftwo
 \else \expandafter \@secondoftwo
 \fi
}%
\providecommand \@ifx [1]{%
 \ifx #1\expandafter \@firstoftwo
 \else \expandafter \@secondoftwo
 \fi
}%
\providecommand \natexlab [1]{#1}%
\providecommand \enquote  [1]{``#1''}%
\providecommand \bibnamefont  [1]{#1}%
\providecommand \bibfnamefont [1]{#1}%
\providecommand \citenamefont [1]{#1}%
\providecommand \href@noop [0]{\@secondoftwo}%
\providecommand \href [0]{\begingroup \@sanitize@url \@href}%
\providecommand \@href[1]{\@@startlink{#1}\@@href}%
\providecommand \@@href[1]{\endgroup#1\@@endlink}%
\providecommand \@sanitize@url [0]{\catcode `\\12\catcode `\$12\catcode
  `\&12\catcode `\#12\catcode `\^12\catcode `\_12\catcode `\%12\relax}%
\providecommand \@@startlink[1]{}%
\providecommand \@@endlink[0]{}%
\providecommand \url  [0]{\begingroup\@sanitize@url \@url }%
\providecommand \@url [1]{\endgroup\@href {#1}{\urlprefix }}%
\providecommand \urlprefix  [0]{URL }%
\providecommand \Eprint [0]{\href }%
\providecommand \doibase [0]{http://dx.doi.org/}%
\providecommand \selectlanguage [0]{\@gobble}%
\providecommand \bibinfo  [0]{\@secondoftwo}%
\providecommand \bibfield  [0]{\@secondoftwo}%
\providecommand \translation [1]{[#1]}%
\providecommand \BibitemOpen [0]{}%
\providecommand \bibitemStop [0]{}%
\providecommand \bibitemNoStop [0]{.\EOS\space}%
\providecommand \EOS [0]{\spacefactor3000\relax}%
\providecommand \BibitemShut  [1]{\csname bibitem#1\endcsname}%
\let\auto@bib@innerbib\@empty
%</preamble>
\bibitem [{\citenamefont {Jeffery}(1922)}]{jeffery1922}%
  \BibitemOpen
  \bibfield  {author} {\bibinfo {author} {\bibfnamefont {G.~B.}\ \bibnamefont
  {Jeffery}},\ }\bibfield  {title} {\enquote {\bibinfo {title} {The motion of
  ellipsoidal particles immersed in a viscous fluid},}\ }\href {\doibase
  10.1098/rspa.1922.0078} {\bibfield  {journal} {\bibinfo  {journal}
  {Proceedings of the Royal Society of London. Series A}\ }\textbf {\bibinfo
  {volume} {102}},\ \bibinfo {pages} {161--179} (\bibinfo {year}
  {1922})}\BibitemShut {NoStop}%
\bibitem [{\citenamefont {Parsa}\ \emph {et~al.}(2012)\citenamefont {Parsa},
  \citenamefont {Calzavarini}, \citenamefont {Toschi},\ and\ \citenamefont
  {Voth}}]{Par12}%
  \BibitemOpen
  \bibfield  {author} {\bibinfo {author} {\bibfnamefont {S.}~\bibnamefont
  {Parsa}}, \bibinfo {author} {\bibfnamefont {E.}~\bibnamefont {Calzavarini}},
  \bibinfo {author} {\bibfnamefont {F.}~\bibnamefont {Toschi}}, \ and\ \bibinfo
  {author} {\bibfnamefont {G.~A.}\ \bibnamefont {Voth}},\ }\bibfield  {title}
  {\enquote {\bibinfo {title} {Rotation rate of rods in turbulent fluid
  flow},}\ }\href@noop {} {\bibfield  {journal} {\bibinfo  {journal} {Phys.
  Rev. Lett.}\ }\textbf {\bibinfo {volume} {109}} (\bibinfo {year} {2012})},\
  \bibinfo {note} {134501}\BibitemShut {NoStop}%
\bibitem [{\citenamefont {Gustavsson}\ \emph
  {et~al.}(2014{\natexlab{a}})\citenamefont {Gustavsson}, \citenamefont
  {Einarsson},\ and\ \citenamefont {Mehlig}}]{Gus14b}%
  \BibitemOpen
  \bibfield  {author} {\bibinfo {author} {\bibfnamefont {K.}~\bibnamefont
  {Gustavsson}}, \bibinfo {author} {\bibfnamefont {J.}~\bibnamefont
  {Einarsson}}, \ and\ \bibinfo {author} {\bibfnamefont {B.}~\bibnamefont
  {Mehlig}},\ }\bibfield  {title} {\enquote {\bibinfo {title} {Tumbling of
  small axisymmetric particles in random and turbulent flows},}\ }\href@noop {}
  {\bibfield  {journal} {\bibinfo  {journal} {Phys. Rev. Lett.}\ }\textbf
  {\bibinfo {volume} {112}},\ \bibinfo {pages} {014501} (\bibinfo {year}
  {2014}{\natexlab{a}})}\BibitemShut {NoStop}%
\bibitem [{\citenamefont {Voth}\ and\ \citenamefont {Soldati}(2016)}]{Voth16}%
  \BibitemOpen
  \bibfield  {author} {\bibinfo {author} {\bibfnamefont {G.}~\bibnamefont
  {Voth}}\ and\ \bibinfo {author} {\bibfnamefont {A.}~\bibnamefont {Soldati}},\
  }\href@noop {} {\bibfield  {journal} {\bibinfo  {journal} {Annu. Rev. Fluid
  Mech.}\ } (\bibinfo {year} {2016})}\BibitemShut {NoStop}%
\bibitem [{\citenamefont {Marchioli}\ \emph {et~al.}(2010)\citenamefont
  {Marchioli}, \citenamefont {Fantoni},\ and\ \citenamefont
  {Soldati}}]{Marchioli2010}%
  \BibitemOpen
  \bibfield  {author} {\bibinfo {author} {\bibfnamefont {C.}~\bibnamefont
  {Marchioli}}, \bibinfo {author} {\bibfnamefont {M.}~\bibnamefont {Fantoni}},
  \ and\ \bibinfo {author} {\bibfnamefont {A.}~\bibnamefont {Soldati}},\
  }\bibfield  {title} {\enquote {\bibinfo {title} {Orientation, distribution,
  and deposition of elongated, inertial fibers in turbulent channel flow},}\
  }\href@noop {} {\bibfield  {journal} {\bibinfo  {journal} {Phys. Fluids}\
  }\textbf {\bibinfo {volume} {22}},\ \bibinfo {pages} {033301} (\bibinfo
  {year} {2010})}\BibitemShut {NoStop}%
\bibitem [{\citenamefont {Challabotla}\ \emph {et~al.}(2015)\citenamefont
  {Challabotla}, \citenamefont {Zhao},\ and\ \citenamefont
  {Andersson}}]{Challabotla2015}%
  \BibitemOpen
  \bibfield  {author} {\bibinfo {author} {\bibfnamefont {N.~R.}\ \bibnamefont
  {Challabotla}}, \bibinfo {author} {\bibfnamefont {L.}~\bibnamefont {Zhao}}, \
  and\ \bibinfo {author} {\bibfnamefont {H.}~\bibnamefont {Andersson}},\
  }\bibfield  {title} {\enquote {\bibinfo {title} {Orientation and rotation of
  inertial disk particles in wall turbulence},}\ }\href@noop {} {\bibfield
  {journal} {\bibinfo  {journal} {J. Fluid Mech.}\ }\textbf {\bibinfo {volume}
  {766}},\ \bibinfo {pages} {R2} (\bibinfo {year} {2015})}\BibitemShut
  {NoStop}%
\bibitem [{\citenamefont {Voth}(2015)}]{Voth15}%
  \BibitemOpen
  \bibfield  {author} {\bibinfo {author} {\bibfnamefont {G.}~\bibnamefont
  {Voth}},\ }\bibfield  {title} {\enquote {\bibinfo {title} {Disks aligned in a
  turbulent channel},}\ }\href@noop {} {\bibfield  {journal} {\bibinfo
  {journal} {J. Fluid Mech.}\ }\textbf {\bibinfo {volume} {772}},\ \bibinfo
  {pages} {1} (\bibinfo {year} {2015})}\BibitemShut {NoStop}%
\bibitem [{\citenamefont {Saffman}(1956)}]{saffman1956}%
  \BibitemOpen
  \bibfield  {author} {\bibinfo {author} {\bibfnamefont {P.~G.}\ \bibnamefont
  {Saffman}},\ }\bibfield  {title} {\enquote {\bibinfo {title} {On the motion
  of small spheroidal particles in a viscous liquid},}\ }\href@noop {}
  {\bibfield  {journal} {\bibinfo  {journal} {J. Fluid Mech.}\ }\textbf
  {\bibinfo {volume} {1}},\ \bibinfo {pages} {540} (\bibinfo {year}
  {1956})}\BibitemShut {NoStop}%
\bibitem [{\citenamefont {Subramanian}\ and\ \citenamefont
  {Koch}(2005)}]{subramanian2005}%
  \BibitemOpen
  \bibfield  {author} {\bibinfo {author} {\bibfnamefont {G.}~\bibnamefont
  {Subramanian}}\ and\ \bibinfo {author} {\bibfnamefont {Donald~L.}\
  \bibnamefont {Koch}},\ }\bibfield  {title} {\enquote {\bibinfo {title}
  {Inertial effects on fibre motion in simple shear flow},}\ }\href {\doibase
  10.1017/S0022112005004829} {\bibfield  {journal} {\bibinfo  {journal}
  {Journal of Fluid Mechanics}\ }\textbf {\bibinfo {volume} {535}},\ \bibinfo
  {pages} {383--414} (\bibinfo {year} {2005})}\BibitemShut {NoStop}%
\bibitem [{\citenamefont {Einarsson}\ \emph
  {et~al.}(2015{\natexlab{a}})\citenamefont {Einarsson}, \citenamefont
  {Candelier}, \citenamefont {Lundell}, \citenamefont {Angilella},\ and\
  \citenamefont {Mehlig}}]{einarsson2015a}%
  \BibitemOpen
  \bibfield  {author} {\bibinfo {author} {\bibfnamefont {J.}~\bibnamefont
  {Einarsson}}, \bibinfo {author} {\bibfnamefont {F.}~\bibnamefont
  {Candelier}}, \bibinfo {author} {\bibfnamefont {F.}~\bibnamefont {Lundell}},
  \bibinfo {author} {\bibfnamefont {J.R.}\ \bibnamefont {Angilella}}, \ and\
  \bibinfo {author} {\bibfnamefont {B.}~\bibnamefont {Mehlig}},\ }\bibfield
  {title} {\enquote {\bibinfo {title} {Rotation of a spheroid in a simple shear
  at small {R}eynolds number},}\ }\href@noop {} {\bibfield  {journal} {\bibinfo
   {journal} {Phys. Fluids}\ }\textbf {\bibinfo {volume} {27}},\ \bibinfo
  {pages} {063301} (\bibinfo {year} {2015}{\natexlab{a}})}\BibitemShut
  {NoStop}%
\bibitem [{\citenamefont {Ros\'{e}n}\ \emph {et~al.}(2015)\citenamefont
  {Ros\'{e}n}, \citenamefont {Einarsson}, \citenamefont {Nordmark},
  \citenamefont {Aidun}, \citenamefont {Lundell},\ and\ \citenamefont
  {Mehlig}}]{rosen2015d}%
  \BibitemOpen
  \bibfield  {author} {\bibinfo {author} {\bibfnamefont {T.}~\bibnamefont
  {Ros\'{e}n}}, \bibinfo {author} {\bibfnamefont {J.}~\bibnamefont
  {Einarsson}}, \bibinfo {author} {\bibfnamefont {A.}~\bibnamefont {Nordmark}},
  \bibinfo {author} {\bibfnamefont {C.~K.}\ \bibnamefont {Aidun}}, \bibinfo
  {author} {\bibfnamefont {F.}~\bibnamefont {Lundell}}, \ and\ \bibinfo
  {author} {\bibfnamefont {B.}~\bibnamefont {Mehlig}},\ }\bibfield  {title}
  {\enquote {\bibinfo {title} {Numerical analysis of the angular motion of a
  neutrally buoyant spheroid in shear flow at small {R}eynolds numbers},}\
  }\href@noop {} {\bibfield  {journal} {\bibinfo  {journal} {Phys. Rev. E}\
  }\textbf {\bibinfo {volume} {92}} (\bibinfo {year} {2015})},\ \bibinfo {note}
  {063022}\BibitemShut {NoStop}%
\bibitem [{\citenamefont {Xu}\ \emph {et~al.}(2011)\citenamefont {Xu},
  \citenamefont {Pumir},\ and\ \citenamefont {Bodenschatz}}]{xu2011}%
  \BibitemOpen
  \bibfield  {author} {\bibinfo {author} {\bibfnamefont {H.}~\bibnamefont
  {Xu}}, \bibinfo {author} {\bibfnamefont {A.}~\bibnamefont {Pumir}}, \ and\
  \bibinfo {author} {\bibfnamefont {E.}~\bibnamefont {Bodenschatz}},\
  }\bibfield  {title} {\enquote {\bibinfo {title} {The pirouette effect in
  turbulent flows},}\ }\href@noop {} {\bibfield  {journal} {\bibinfo  {journal}
  {Nature Physics}\ }\textbf {\bibinfo {volume} {7}},\ \bibinfo {pages} {709}
  (\bibinfo {year} {2011})}\BibitemShut {NoStop}%
\bibitem [{\citenamefont {Jucha}\ \emph {et~al.}(2014)\citenamefont {Jucha},
  \citenamefont {Xu}, \citenamefont {Pumir},\ and\ \citenamefont
  {Bodenschatz}}]{jucha2014}%
  \BibitemOpen
  \bibfield  {author} {\bibinfo {author} {\bibfnamefont {J.}~\bibnamefont
  {Jucha}}, \bibinfo {author} {\bibfnamefont {H.}~\bibnamefont {Xu}}, \bibinfo
  {author} {\bibfnamefont {A.}~\bibnamefont {Pumir}}, \ and\ \bibinfo {author}
  {\bibfnamefont {E}~\bibnamefont {Bodenschatz}},\ }\bibfield  {title}
  {\enquote {\bibinfo {title} {Time-reversal-symmetry breaking in
  turbulence},}\ }\href@noop {} {\bibfield  {journal} {\bibinfo  {journal}
  {Phys. Rev. Lett.}\ }\textbf {\bibinfo {volume} {113}},\ \bibinfo {pages}
  {054501} (\bibinfo {year} {2014})}\BibitemShut {NoStop}%
\bibitem [{\citenamefont {Xu}\ \emph {et~al.}(2014)\citenamefont {Xu},
  \citenamefont {Pumir}, \citenamefont {Falkovich}, \citenamefont
  {Bodenschatz}, \citenamefont {Shats}, \citenamefont {Xia}, \citenamefont
  {Francois},\ and\ \citenamefont {Boffetta}}]{xu2014}%
  \BibitemOpen
  \bibfield  {author} {\bibinfo {author} {\bibfnamefont {H.}~\bibnamefont
  {Xu}}, \bibinfo {author} {\bibfnamefont {A.}~\bibnamefont {Pumir}}, \bibinfo
  {author} {\bibfnamefont {G.}~\bibnamefont {Falkovich}}, \bibinfo {author}
  {\bibfnamefont {E.}~\bibnamefont {Bodenschatz}}, \bibinfo {author}
  {\bibfnamefont {M.}~\bibnamefont {Shats}}, \bibinfo {author} {\bibfnamefont
  {H.}~\bibnamefont {Xia}}, \bibinfo {author} {\bibfnamefont {N.}~\bibnamefont
  {Francois}}, \ and\ \bibinfo {author} {\bibfnamefont {G.}~\bibnamefont
  {Boffetta}},\ }\bibfield  {title} {\enquote {\bibinfo {title} {Flight-crash
  events in turbulence},}\ }\href@noop {} {\bibfield  {journal} {\bibinfo
  {journal} {Proc. Natl. Acad. Sci. USA}\ }\textbf {\bibinfo {volume} {111}},\
  \bibinfo {pages} {7558} (\bibinfo {year} {2014})}\BibitemShut {NoStop}%
\bibitem [{\citenamefont {Pumir}\ \emph {et~al.}(2016)\citenamefont {Pumir},
  \citenamefont {Xu}, \citenamefont {Bodenschatz},\ and\ \citenamefont
  {Grauer}}]{pumir2016}%
  \BibitemOpen
  \bibfield  {author} {\bibinfo {author} {\bibfnamefont {A.}~\bibnamefont
  {Pumir}}, \bibinfo {author} {\bibfnamefont {H.}~\bibnamefont {Xu}}, \bibinfo
  {author} {\bibfnamefont {E.}~\bibnamefont {Bodenschatz}}, \ and\ \bibinfo
  {author} {\bibfnamefont {R.}~\bibnamefont {Grauer}},\ }\bibfield  {title}
  {\enquote {\bibinfo {title} {Single-particle motion and vortex stretching in
  three-dimensional turbulent flows},}\ }\href@noop {} {\bibfield  {journal}
  {\bibinfo  {journal} {Phys. Rev. Lett.}\ }\textbf {\bibinfo {volume} {16}},\
  \bibinfo {pages} {124502} (\bibinfo {year} {2016})}\BibitemShut {NoStop}%
\bibitem [{\citenamefont {Lovalenti}\ and\ \citenamefont
  {Brady}(1993)}]{lovalenti1993}%
  \BibitemOpen
  \bibfield  {author} {\bibinfo {author} {\bibfnamefont {P.M.}\ \bibnamefont
  {Lovalenti}}\ and\ \bibinfo {author} {\bibfnamefont {J.F.}\ \bibnamefont
  {Brady}},\ }\bibfield  {title} {\enquote {\bibinfo {title} {The force on a
  bubble, drop or particle in arbitrary time-dependent motion at small
  {R}eynolds number.}}\ }\href@noop {} {\bibfield  {journal} {\bibinfo
  {journal} {Phys. Fluids}\ }\textbf {\bibinfo {volume} {5}},\ \bibinfo {pages}
  {2104--2116} (\bibinfo {year} {1993})}\BibitemShut {NoStop}%
\bibitem [{\citenamefont {Einarsson}\ \emph
  {et~al.}(2015{\natexlab{b}})\citenamefont {Einarsson}, \citenamefont
  {Candelier}, \citenamefont {Lundell}, \citenamefont {Angilella},\ and\
  \citenamefont {Mehlig}}]{einarsson2015b}%
  \BibitemOpen
  \bibfield  {author} {\bibinfo {author} {\bibfnamefont {J.}~\bibnamefont
  {Einarsson}}, \bibinfo {author} {\bibfnamefont {F.}~\bibnamefont
  {Candelier}}, \bibinfo {author} {\bibfnamefont {F.}~\bibnamefont {Lundell}},
  \bibinfo {author} {\bibfnamefont {J.R.}\ \bibnamefont {Angilella}}, \ and\
  \bibinfo {author} {\bibfnamefont {B.}~\bibnamefont {Mehlig}},\ }\bibfield
  {title} {\enquote {\bibinfo {title} {Effect of weak fluid inertia upon
  {Jeffery} orbits},}\ }\href@noop {} {\bibfield  {journal} {\bibinfo
  {journal} {Phys. Rev. E}\ }\textbf {\bibinfo {volume} {91}},\ \bibinfo
  {pages} {041002(R)} (\bibinfo {year} {2015}{\natexlab{b}})}\BibitemShut
  {NoStop}%
\bibitem [{\citenamefont {Candelier}\ \emph {et~al.}(2015)\citenamefont
  {Candelier}, \citenamefont {Einarsson}, \citenamefont {Lundell},
  \citenamefont {Mehlig},\ and\ \citenamefont {Angilella}}]{candelier2015b}%
  \BibitemOpen
  \bibfield  {author} {\bibinfo {author} {\bibfnamefont {F.}~\bibnamefont
  {Candelier}}, \bibinfo {author} {\bibfnamefont {J.}~\bibnamefont
  {Einarsson}}, \bibinfo {author} {\bibfnamefont {F.}~\bibnamefont {Lundell}},
  \bibinfo {author} {\bibfnamefont {B.}~\bibnamefont {Mehlig}}, \ and\ \bibinfo
  {author} {\bibfnamefont {J.R.}\ \bibnamefont {Angilella}},\ }\bibfield
  {title} {\enquote {\bibinfo {title} {The role of inertia for the rotation of
  a nearly spherical particle in a general linear flow},}\ }\href@noop {}
  {\bibfield  {journal} {\bibinfo  {journal} {Phys. Rev. E}\ }\textbf {\bibinfo
  {volume} {91}},\ \bibinfo {pages} {053023\,(\mbox{erratum}\, 059901)}
  (\bibinfo {year} {2015})}\BibitemShut {NoStop}%
\bibitem [{\citenamefont {Kim}\ and\ \citenamefont {Karrila}(1991)}]{kim1991}%
  \BibitemOpen
  \bibfield  {author} {\bibinfo {author} {\bibfnamefont {Sangtae}\ \bibnamefont
  {Kim}}\ and\ \bibinfo {author} {\bibfnamefont {Seppo~J.}\ \bibnamefont
  {Karrila}},\ }\href@noop {} {\emph {\bibinfo {title} {Microhydrodynamics:
  principles and selected applications}}}\ (\bibinfo  {publisher}
  {Butterworth-Heinemann},\ \bibinfo {address} {Boston},\ \bibinfo {year}
  {1991})\BibitemShut {NoStop}%
\bibitem [{\citenamefont {{}}(2016)}]{supp}%
  \BibitemOpen
  \bibfield  {author} {\bibinfo {author} {\bibnamefont {{ See Supplemental Material [url], which includes Ref.~[21]}}}.}
%  \bibfield  {title} {\enquote {\bibinfo {title}
%  {}}\ }\href@noop {} {\  (\bibinfo
%  {year} {2016})}
\BibitemShut {NoStop}%
\bibitem [{\citenamefont {Happel}\ and\ \citenamefont
  {Brenner}(1983)}]{happel1983}%
  \BibitemOpen
  \bibfield  {author} {\bibinfo {author} {\bibfnamefont {J.}~\bibnamefont
  {Happel}}\ and\ \bibinfo {author} {\bibfnamefont {H.}~\bibnamefont
  {Brenner}},\ }\href@noop {} {\emph {\bibinfo {title} {Low {R}eynolds number
  hydrodynamics}}}\ (\bibinfo  {publisher} {Kluwer Acad. Publisher},\ \bibinfo
  {year} {1983})\BibitemShut {NoStop}%
\bibitem [{\citenamefont {Maxey}(1987)}]{Max87}%
  \BibitemOpen
  \bibfield  {author} {\bibinfo {author} {\bibfnamefont {M.~R.}\ \bibnamefont
  {Maxey}},\ }\bibfield  {title} {\enquote {\bibinfo {title} {The gravitational
  settling of aerosol particles in homogeneous turbulence and random flow
  fields},}\ }\href@noop {} {\bibfield  {journal} {\bibinfo  {journal} {J.
  Fluid Mech.}\ }\textbf {\bibinfo {volume} {174}},\ \bibinfo {pages}
  {441--465} (\bibinfo {year} {1987})}\BibitemShut {NoStop}%
\bibitem [{\citenamefont {Gustavson}\ and\ \citenamefont
  {Mehlig}(2016)}]{Gus16}%
  \BibitemOpen
  \bibfield  {author} {\bibinfo {author} {\bibfnamefont {K.}~\bibnamefont
  {Gustavson}}\ and\ \bibinfo {author} {\bibfnamefont {B.}~\bibnamefont
  {Mehlig}},\ }\bibfield  {title} {\enquote {\bibinfo {title} {Statistical
  models for spatial patterns of heavy particles in turbulence},}\ }\href@noop
  {} {\bibfield  {journal} {\bibinfo  {journal} {Adv. Phys.}\ }\textbf
  {\bibinfo {volume} {65}},\ \bibinfo {pages} {1} (\bibinfo {year}
  {2016})}\BibitemShut {NoStop}%
\bibitem [{\citenamefont {Lin}\ \emph {et~al.}(1970)\citenamefont {Lin},
  \citenamefont {Peery},\ and\ \citenamefont {Schowalter}}]{lin1970}%
  \BibitemOpen
  \bibfield  {author} {\bibinfo {author} {\bibfnamefont {C.~J.}\ \bibnamefont
  {Lin}}, \bibinfo {author} {\bibfnamefont {J.~H.}\ \bibnamefont {Peery}}, \
  and\ \bibinfo {author} {\bibfnamefont {W.~R.}\ \bibnamefont {Schowalter}},\
  }\bibfield  {title} {\enquote {\bibinfo {title} {Simple shear flow around a
  rigid sphere: inertial effects and suspension rheology},}\ }\href@noop {}
  {\bibfield  {journal} {\bibinfo  {journal} {J. Fluid Mech.}\ }\textbf
  {\bibinfo {volume} {44}},\ \bibinfo {pages} {1} (\bibinfo {year}
  {1970})}\BibitemShut {NoStop}%
\bibitem [{\citenamefont {Meibohm}\ \emph {et~al.}(2016)\citenamefont
  {Meibohm}, \citenamefont {Candelier}, \citenamefont {Rosen}, \citenamefont
  {Einarsson}, \citenamefont {Lundell},\ and\ \citenamefont
  {Mehlig}}]{meibohm2016}%
  \BibitemOpen
  \bibfield  {author} {\bibinfo {author} {\bibfnamefont {J.}~\bibnamefont
  {Meibohm}}, \bibinfo {author} {\bibfnamefont {F.}~\bibnamefont {Candelier}},
  \bibinfo {author} {\bibfnamefont {T.}~\bibnamefont {Rosen}}, \bibinfo
  {author} {\bibfnamefont {J.}~\bibnamefont {Einarsson}}, \bibinfo {author}
  {\bibfnamefont {F.}~\bibnamefont {Lundell}}, \ and\ \bibinfo {author}
  {\bibfnamefont {B.}~\bibnamefont {Mehlig}},\ }\bibfield  {title} {\enquote
  {\bibinfo {title} {Angular velocity of a sphere in a simple shear at small
  {R}eynolds number},}\ }\href@noop {} {\  (\bibinfo {year} {2016})},\ \bibinfo
  {note} {{submitted to Phys. Rev. Fluids}}\BibitemShut {NoStop}%
\bibitem [{\citenamefont {Li}\ \emph {et~al.}(2008)\citenamefont {Li},
  \citenamefont {Perlman}, \citenamefont {Wan}, \citenamefont {Yang},
  \citenamefont {Meneveau}, \citenamefont {Burns}, \citenamefont {Chen},
  \citenamefont {Szalay},\ and\ \citenamefont {Eyink}}]{Yi2008}%
  \BibitemOpen
  \bibfield  {author} {\bibinfo {author} {\bibfnamefont {Yi}~\bibnamefont
  {Li}}, \bibinfo {author} {\bibfnamefont {Eric}\ \bibnamefont {Perlman}},
  \bibinfo {author} {\bibfnamefont {Minping}\ \bibnamefont {Wan}}, \bibinfo
  {author} {\bibfnamefont {Yunke}\ \bibnamefont {Yang}}, \bibinfo {author}
  {\bibfnamefont {Charles}\ \bibnamefont {Meneveau}}, \bibinfo {author}
  {\bibfnamefont {Randal}\ \bibnamefont {Burns}}, \bibinfo {author}
  {\bibfnamefont {Shiyi}\ \bibnamefont {Chen}}, \bibinfo {author}
  {\bibfnamefont {Alexander}\ \bibnamefont {Szalay}}, \ and\ \bibinfo {author}
  {\bibfnamefont {Gregory}\ \bibnamefont {Eyink}},\ }\bibfield  {title}
  {\enquote {\bibinfo {title} {A public turbulence database cluster and
  applications to study lagrangian evolution of velocity increments in
  turbulence},}\ }\href@noop {} {\bibfield  {journal} {\bibinfo  {journal}
  {Journal of Turbulence}\ } (\bibinfo {year} {2008})},\ \bibinfo {note}
  {n31}\BibitemShut {NoStop}%
\bibitem [{\citenamefont {Yu}\ \emph {et~al.}(2012)\citenamefont {Yu},
  \citenamefont {Kanov}, \citenamefont {Perlman}, \citenamefont {Graham},
  \citenamefont {Frederix}, \citenamefont {Burns}, \citenamefont {Szalay},
  \citenamefont {Eyink},\ and\ \citenamefont {Meneveau}}]{Yu2012}%
  \BibitemOpen
  \bibfield  {author} {\bibinfo {author} {\bibfnamefont {H.}~\bibnamefont
  {Yu}}, \bibinfo {author} {\bibfnamefont {K.}~\bibnamefont {Kanov}}, \bibinfo
  {author} {\bibfnamefont {E.}~\bibnamefont {Perlman}}, \bibinfo {author}
  {\bibfnamefont {J.}~\bibnamefont {Graham}}, \bibinfo {author} {\bibfnamefont
  {E.}~\bibnamefont {Frederix}}, \bibinfo {author} {\bibfnamefont
  {R.}~\bibnamefont {Burns}}, \bibinfo {author} {\bibfnamefont
  {A.}~\bibnamefont {Szalay}}, \bibinfo {author} {\bibfnamefont
  {G.}~\bibnamefont {Eyink}}, \ and\ \bibinfo {author} {\bibfnamefont
  {C.}~\bibnamefont {Meneveau}},\ }\bibfield  {title} {\enquote {\bibinfo
  {title} {A public turbulence database cluster and applications to study
  lagrangian evolution of velocity increments in turbulence},}\ }\href@noop {}
  {\bibfield  {journal} {\bibinfo  {journal} {Journal of Turbulence}\ }
  (\bibinfo {year} {2012})}\BibitemShut {NoStop}%
\bibitem [{\citenamefont {Guala}\ \emph {et~al.}(2005)\citenamefont {Guala},
  \citenamefont {{L\"uthi}}, \citenamefont {Liberzon}, \citenamefont
  {Tsinober},\ and\ \citenamefont {Kinzelbach}}]{guala2005}%
  \BibitemOpen
  \bibfield  {author} {\bibinfo {author} {\bibfnamefont {M.}~\bibnamefont
  {Guala}}, \bibinfo {author} {\bibfnamefont {B.}~\bibnamefont {{L\"uthi}}},
  \bibinfo {author} {\bibfnamefont {A.}~\bibnamefont {Liberzon}}, \bibinfo
  {author} {\bibfnamefont {A.}~\bibnamefont {Tsinober}}, \ and\ \bibinfo
  {author} {\bibfnamefont {W.}~\bibnamefont {Kinzelbach}},\ }\bibfield  {title}
  {\enquote {\bibinfo {title} {On the evolution of material lines and vorticity
  in homogeneous turbulence},}\ }\href@noop {} {\bibfield  {journal} {\bibinfo
  {journal} {J. Fluid Mech.}\ }\textbf {\bibinfo {volume} {533}},\ \bibinfo
  {pages} {339} (\bibinfo {year} {2005})}\BibitemShut {NoStop}%
\bibitem [{\citenamefont {{L\"uthi}}\ \emph {et~al.}(2005)\citenamefont
  {{L\"uthi}}, \citenamefont {Tsinober},\ and\ \citenamefont
  {Kinzelbach}}]{luthi2005}%
  \BibitemOpen
  \bibfield  {author} {\bibinfo {author} {\bibfnamefont {B.}~\bibnamefont
  {{L\"uthi}}}, \bibinfo {author} {\bibfnamefont {A.}~\bibnamefont {Tsinober}},
  \ and\ \bibinfo {author} {\bibfnamefont {W.}~\bibnamefont {Kinzelbach}},\
  }\bibfield  {title} {\enquote {\bibinfo {title} {Lagrangian measurement of
  vorticity dynamics in turbulent flow},}\ }\href@noop {} {\bibfield  {journal}
  {\bibinfo  {journal} {J. Fluid Mech.}\ }\textbf {\bibinfo {volume} {528}},\
  \bibinfo {pages} {87} (\bibinfo {year} {2005})}\BibitemShut {NoStop}%
\bibitem [{\citenamefont {Johnson}\ and\ \citenamefont
  {Meneveau}(2016)}]{johnson2016}%
  \BibitemOpen
  \bibfield  {author} {\bibinfo {author} {\bibfnamefont {P.~L.}\ \bibnamefont
  {Johnson}}\ and\ \bibinfo {author} {\bibfnamefont {C.}~\bibnamefont
  {Meneveau}},\ }\bibfield  {title} {\enquote {\bibinfo {title}
  {Large-deviation statistics of vorticity stretching in isotropic
  turbulence},}\ }\href@noop {} {\bibfield  {journal} {\bibinfo  {journal}
  {Phys. Rev. E}\ }\textbf {\bibinfo {volume} {93}},\ \bibinfo {pages} {033118}
  (\bibinfo {year} {2016})}\BibitemShut {NoStop}%
\bibitem [{\citenamefont {Schumacher}\ \emph {et~al.}(2014)\citenamefont
  {Schumacher}, \citenamefont {Scheel}, \citenamefont {Krasnov}, \citenamefont
  {Donzis}, \citenamefont {Yakhot},\ and\ \citenamefont
  {Sreenivasan}}]{schumacher2014}%
  \BibitemOpen
  \bibfield  {author} {\bibinfo {author} {\bibfnamefont {J.}~\bibnamefont
  {Schumacher}}, \bibinfo {author} {\bibfnamefont {J.~D.}\ \bibnamefont
  {Scheel}}, \bibinfo {author} {\bibfnamefont {D.}~\bibnamefont {Krasnov}},
  \bibinfo {author} {\bibfnamefont {D.~A.}\ \bibnamefont {Donzis}}, \bibinfo
  {author} {\bibfnamefont {V.}~\bibnamefont {Yakhot}}, \ and\ \bibinfo {author}
  {\bibfnamefont {K.~R.}\ \bibnamefont {Sreenivasan}},\ }\bibfield  {title}
  {\enquote {\bibinfo {title} {Small-scale universality in fluid turbulence},}\
  }\href@noop {} {\bibfield  {journal} {\bibinfo  {journal} {PNAS}\ }\textbf
  {\bibinfo {volume} {111}},\ \bibinfo {pages} {10961--10965} (\bibinfo {year}
  {2014})}\BibitemShut {NoStop}%
\bibitem [{\citenamefont {Byron}\ \emph {et~al.}(2015)\citenamefont {Byron},
  \citenamefont {Einarsson}, \citenamefont {Gustavsson}, \citenamefont {Voth},
  \citenamefont {Mehlig},\ and\ \citenamefont {Variano}}]{Byron2015}%
  \BibitemOpen
  \bibfield  {author} {\bibinfo {author} {\bibfnamefont {M.}~\bibnamefont
  {Byron}}, \bibinfo {author} {\bibfnamefont {J.}~\bibnamefont {Einarsson}},
  \bibinfo {author} {\bibfnamefont {K.}~\bibnamefont {Gustavsson}}, \bibinfo
  {author} {\bibfnamefont {G.}~\bibnamefont {Voth}}, \bibinfo {author}
  {\bibfnamefont {B.}~\bibnamefont {Mehlig}}, \ and\ \bibinfo {author}
  {\bibfnamefont {E.}~\bibnamefont {Variano}},\ }\bibfield  {title} {\enquote
  {\bibinfo {title} {Shape-dependence of particle rotation in isotropic
  turbulence},}\ }\href@noop {} {\bibfield  {journal} {\bibinfo  {journal}
  {Phys. Fluids}\ }\textbf {\bibinfo {volume} {27}},\ \bibinfo {pages} {035101}
  (\bibinfo {year} {2015})}\BibitemShut {NoStop}%
\bibitem [{\citenamefont {Gatignol}(1983)}]{gatignol1983}%
  \BibitemOpen
  \bibfield  {author} {\bibinfo {author} {\bibfnamefont {R.}~\bibnamefont
  {Gatignol}},\ }\bibfield  {title} {\enquote {\bibinfo {title} {The {F}ax\'en
  formulae for a rigid particle in an unsteady non-uniform {S}tokes flow.}}\
  }\href@noop {} {\bibfield  {journal} {\bibinfo  {journal} {J. M\'ec. Th\'eor.
  Appl.}\ }\textbf {\bibinfo {volume} {1}},\ \bibinfo {pages} {143--160}
  (\bibinfo {year} {1983})}\BibitemShut {NoStop}%
\bibitem [{\citenamefont {Maxey}\ and\ \citenamefont
  {Riley}(1983)}]{maxey1983}%
  \BibitemOpen
  \bibfield  {author} {\bibinfo {author} {\bibfnamefont {M.~R.}\ \bibnamefont
  {Maxey}}\ and\ \bibinfo {author} {\bibfnamefont {J.~J.}\ \bibnamefont
  {Riley}},\ }\bibfield  {title} {\enquote {\bibinfo {title} {Equation of
  motion for a small rigid sphere in a nonuniform flow},}\ }\href@noop {}
  {\bibfield  {journal} {\bibinfo  {journal} {Phys. Fluids}\ }\textbf {\bibinfo
  {volume} {26}},\ \bibinfo {pages} {883--889} (\bibinfo {year}
  {1983})}\BibitemShut {NoStop}%
\bibitem [{\citenamefont {Traugott}\ and\ \citenamefont
  {Liberzon}(2015)}]{Tra15}%
  \BibitemOpen
  \bibfield  {author} {\bibinfo {author} {\bibfnamefont {H.}~\bibnamefont
  {Traugott}}\ and\ \bibinfo {author} {\bibfnamefont {A.}~\bibnamefont
  {Liberzon}},\ }\bibfield  {title} {\enquote {\bibinfo {title} {Experimental
  study of forces on freely moving spherical particles during resuspension into
  turbulent flow},}\ }\href@noop {} {\bibfield  {journal} {\bibinfo  {journal}
  {arXiv:1510.00879}\ } (\bibinfo {year} {2015})}\BibitemShut {NoStop}%
\bibitem [{\citenamefont {Variano}(2016)}]{variano2016}%
  \BibitemOpen
  \bibfield  {author} {\bibinfo {author} {\bibfnamefont {E.}~\bibnamefont
  {Variano}},\ }\bibfield  {title} {\enquote {\bibinfo {title} {private
  communication},}\ }\href@noop {} {\  (\bibinfo {year} {2016})}\BibitemShut
  {NoStop}%
\bibitem [{\citenamefont {Marcus}\ \emph {et~al.}(2014)\citenamefont {Marcus},
  \citenamefont {Parsa}, \citenamefont {Kramel}, \citenamefont {Ni},\ and\
  \citenamefont {Voth}}]{Voth}%
  \BibitemOpen
  \bibfield  {author} {\bibinfo {author} {\bibfnamefont {G.~G.}\ \bibnamefont
  {Marcus}}, \bibinfo {author} {\bibfnamefont {S.}~\bibnamefont {Parsa}},
  \bibinfo {author} {\bibfnamefont {S.}~\bibnamefont {Kramel}}, \bibinfo
  {author} {\bibfnamefont {R.}~\bibnamefont {Ni}}, \ and\ \bibinfo {author}
  {\bibfnamefont {G.~A.}\ \bibnamefont {Voth}},\ }\bibfield  {title} {\enquote
  {\bibinfo {title} {Measurements of the solid-body rotation of anisotropic
  particles in 3d turbulence},}\ }\href@noop {} {\bibfield  {journal} {\bibinfo
   {journal} {New J. Phys.}\ }\textbf {\bibinfo {volume} {16}},\ \bibinfo
  {pages} {102001} (\bibinfo {year} {2014})}\BibitemShut {NoStop}%
\bibitem [{\citenamefont {Homann}\ and\ \citenamefont {Bec}(2010)}]{Homann}%
  \BibitemOpen
  \bibfield  {author} {\bibinfo {author} {\bibfnamefont {H.}~\bibnamefont
  {Homann}}\ and\ \bibinfo {author} {\bibfnamefont {J.}~\bibnamefont {Bec}},\
  }\bibfield  {title} {\enquote {\bibinfo {title} {Finite-size effects in the
  dynamics of neutrally buoyant particles in turbulent flow},}\ }\href@noop {}
  {\bibfield  {journal} {\bibinfo  {journal} {J. Fluid Mech.}\ }\textbf
  {\bibinfo {volume} {651}},\ \bibinfo {pages} {81--91} (\bibinfo {year}
  {2010})}\BibitemShut {NoStop}%
\bibitem [{\citenamefont {Fornari}\ \emph {et~al.}(2016)\citenamefont
  {Fornari}, \citenamefont {Picano}, \citenamefont {Sardina},\ and\
  \citenamefont {Brandt}}]{Fornari}%
  \BibitemOpen
  \bibfield  {author} {\bibinfo {author} {\bibfnamefont {W.}~\bibnamefont
  {Fornari}}, \bibinfo {author} {\bibfnamefont {F.}~\bibnamefont {Picano}},
  \bibinfo {author} {\bibfnamefont {G.}~\bibnamefont {Sardina}}, \ and\
  \bibinfo {author} {\bibfnamefont {L.}~\bibnamefont {Brandt}},\ }\bibfield
  {title} {\enquote {\bibinfo {title} {Reduced particle settling speed in
  turbulence},}\ }\href@noop {} {\bibfield  {journal} {\bibinfo  {journal}
  {submitted to J. Fluid Mech.}\ } (\bibinfo {year} {2016})}\BibitemShut
  {NoStop}%
\bibitem [{\citenamefont {Gustavsson}\ \emph
  {et~al.}(2014{\natexlab{b}})\citenamefont {Gustavsson}, \citenamefont
  {Vajedi},\ and\ \citenamefont {Mehlig}}]{Gus14e}%
  \BibitemOpen
  \bibfield  {author} {\bibinfo {author} {\bibfnamefont {K.}~\bibnamefont
  {Gustavsson}}, \bibinfo {author} {\bibfnamefont {S.}~\bibnamefont {Vajedi}},
  \ and\ \bibinfo {author} {\bibfnamefont {B.}~\bibnamefont {Mehlig}},\
  }\bibfield  {title} {\enquote {\bibinfo {title} {Clustering of particles
  falling in a turbulent flow},}\ }\href@noop {} {\bibfield  {journal}
  {\bibinfo  {journal} {Phys. Rev. Lett.}\ }\textbf {\bibinfo {volume} {112}},\
  \bibinfo {pages} {214501} (\bibinfo {year} {2014}{\natexlab{b}})}\BibitemShut
  {NoStop}%
\bibitem [{\citenamefont {Ireland}\ \emph {et~al.}(2016)\citenamefont
  {Ireland}, \citenamefont {Bragg},\ and\ \citenamefont {Collins}}]{Ireland}%
  \BibitemOpen
  \bibfield  {author} {\bibinfo {author} {\bibfnamefont {P.J.}\ \bibnamefont
  {Ireland}}, \bibinfo {author} {\bibfnamefont {A.D.}\ \bibnamefont {Bragg}}, \
  and\ \bibinfo {author} {\bibfnamefont {L.R.}\ \bibnamefont {Collins}},\
  }\bibfield  {title} {\enquote {\bibinfo {title} {The effect of {R}eynolds
  number on inertial particle dynamics in isotropic turbulence. {P}art 2.
  {S}imulations with gravitational effects},}\ }\href@noop {} {\bibfield
  {journal} {\bibinfo  {journal} {J. Fluid Mech.}\ }\textbf {\bibinfo {volume}
  {796}},\ \bibinfo {pages} {659--711} (\bibinfo {year} {2016})}\BibitemShut
  {NoStop}%
\bibitem [{\citenamefont {Mathai}\ \emph {et~al.}(2016)\citenamefont {Mathai},
  \citenamefont {Calzavarini}, \citenamefont {Brons}, \citenamefont {Sun},\
  and\ \citenamefont {Lohse}}]{Mathai2016}%
  \BibitemOpen
  \bibfield  {author} {\bibinfo {author} {\bibfnamefont {V.}~\bibnamefont
  {Mathai}}, \bibinfo {author} {\bibfnamefont {E.}~\bibnamefont {Calzavarini}},
  \bibinfo {author} {\bibfnamefont {J.}~\bibnamefont {Brons}}, \bibinfo
  {author} {\bibfnamefont {C.}~\bibnamefont {Sun}}, \ and\ \bibinfo {author}
  {\bibfnamefont {D.}~\bibnamefont {Lohse}},\ }\bibfield  {title} {\enquote
  {\bibinfo {title} {Microbubbles and microparticles are not faithful tracers
  of turbulent acceleration},}\ }\href@noop {} {\bibfield  {journal} {\bibinfo
  {journal} {Phys. Rev. Lett.}\ }\textbf {\bibinfo {volume} {117}},\ \bibinfo
  {pages} {024501} (\bibinfo {year} {2016})}\BibitemShut {NoStop}%
\bibitem [{\citenamefont {Parishani}\ \emph {et~al.}(2015)\citenamefont
  {Parishani}, \citenamefont {Ayala}, \citenamefont {Rosa}, \citenamefont
  {Wang},\ and\ \citenamefont {Grabowski}}]{Parishani}%
  \BibitemOpen
  \bibfield  {author} {\bibinfo {author} {\bibfnamefont {H.}~\bibnamefont
  {Parishani}}, \bibinfo {author} {\bibfnamefont {O.}~\bibnamefont {Ayala}},
  \bibinfo {author} {\bibfnamefont {B.}~\bibnamefont {Rosa}}, \bibinfo {author}
  {\bibfnamefont {L.-P.}\ \bibnamefont {Wang}}, \ and\ \bibinfo {author}
  {\bibfnamefont {W.~W.}\ \bibnamefont {Grabowski}},\ }\bibfield  {title}
  {\enquote {\bibinfo {title} {Effects of gravity on the acceleration and pair
  statistics of inertial particles in homogeneous isotropic turbulence},}\
  }\href@noop {} {\bibfield  {journal} {\bibinfo  {journal} {Phys. Fluids}\
  }\textbf {\bibinfo {volume} {27}},\ \bibinfo {pages} {033304} (\bibinfo
  {year} {2015})}\BibitemShut {NoStop}%
\bibitem [{\citenamefont {Pruppacher}\ and\ \citenamefont
  {Klett}(1997)}]{Pru78}%
  \BibitemOpen
  \bibfield  {author} {\bibinfo {author} {\bibfnamefont {H.~R.}\ \bibnamefont
  {Pruppacher}}\ and\ \bibinfo {author} {\bibfnamefont {J.~D.}\ \bibnamefont
  {Klett}},\ }\href@noop {} {\emph {\bibinfo {title} {Microphysics of clouds
  and precipitation, 2nd edition}}}\ (\bibinfo  {publisher} {Kluwer Academic
  Publishers},\ \bibinfo {address} {Dordrecht, The Nederlands},\ \bibinfo
  {year} {1997})\ \bibinfo {note} {954p}\BibitemShut {NoStop}%
\bibitem [{\citenamefont {Zimmermann}\ \emph {et~al.}(2011)\citenamefont
  {Zimmermann}, \citenamefont {Gaseuil}, \citenamefont {Bourgoin},
  \citenamefont {R.}, \citenamefont {Pumir},\ and\ \citenamefont
  {Pinton}}]{zimmermann2011}%
  \BibitemOpen
  \bibfield  {author} {\bibinfo {author} {\bibfnamefont {R.}~\bibnamefont
  {Zimmermann}}, \bibinfo {author} {\bibfnamefont {Y.}~\bibnamefont {Gaseuil}},
  \bibinfo {author} {\bibfnamefont {M.}~\bibnamefont {Bourgoin}}, \bibinfo
  {author} {\bibfnamefont {Volk.}\ \bibnamefont {R.}}, \bibinfo {author}
  {\bibfnamefont {A.}~\bibnamefont {Pumir}}, \ and\ \bibinfo {author}
  {\bibfnamefont {J.-F.}\ \bibnamefont {Pinton}},\ }\bibfield  {title}
  {\enquote {\bibinfo {title} {Rotational intermittency and turbulence induced
  lift experienced by large particles in a turbulent flow},}\ }\href@noop {}
  {\bibfield  {journal} {\bibinfo  {journal} {Phys. Rev. Lett.}\ }\textbf
  {\bibinfo {volume} {106}},\ \bibinfo {pages} {154501} (\bibinfo {year}
  {2011})}\BibitemShut {NoStop}%
\bibitem [{\citenamefont {Saffman}(1965)}]{saffman1965}%
  \BibitemOpen
  \bibfield  {author} {\bibinfo {author} {\bibfnamefont {P.~G.}\ \bibnamefont
  {Saffman}},\ }\bibfield  {title} {\enquote {\bibinfo {title} {The lift on a
  small sphere in a slow shear flow},}\ }\href@noop {} {\bibfield  {journal}
  {\bibinfo  {journal} {J. Fluid Mech.}\ }\textbf {\bibinfo {volume} {22}},\
  \bibinfo {pages} {385--400} (\bibinfo {year} {1965})}\BibitemShut {NoStop}%
\bibitem [{\citenamefont {Parsa}\ and\ \citenamefont {Voth}(2014)}]{Par14}%
  \BibitemOpen
  \bibfield  {author} {\bibinfo {author} {\bibfnamefont {S.}~\bibnamefont
  {Parsa}}\ and\ \bibinfo {author} {\bibfnamefont {G.~A.}\ \bibnamefont
  {Voth}},\ }\bibfield  {title} {\enquote {\bibinfo {title} {Inertial range
  scaling in rotations of long rods in turbulence},}\ }\href@noop {} {\bibfield
   {journal} {\bibinfo  {journal} {Phys. Rev. Lett.}\ }\textbf {\bibinfo
  {volume} {112}},\ \bibinfo {pages} {024501} (\bibinfo {year}
  {2014})}\BibitemShut {NoStop}%
\bibitem [{\citenamefont {Mathai}\ \emph {et~al.}(2015)\citenamefont {Mathai},
  \citenamefont {Prakash}, \citenamefont {Brons}, \citenamefont {Sun},\ and\
  \citenamefont {Lohse}}]{Mat15}%
  \BibitemOpen
  \bibfield  {author} {\bibinfo {author} {\bibfnamefont {Varghese}\
  \bibnamefont {Mathai}}, \bibinfo {author} {\bibfnamefont {Vivek~N.}\
  \bibnamefont {Prakash}}, \bibinfo {author} {\bibfnamefont {Jon}\ \bibnamefont
  {Brons}}, \bibinfo {author} {\bibfnamefont {Chao}\ \bibnamefont {Sun}}, \
  and\ \bibinfo {author} {\bibfnamefont {Detlef}\ \bibnamefont {Lohse}},\
  }\bibfield  {title} {\enquote {\bibinfo {title} {Wake-driven dynamics of
  finite-sized buoyant spheres in turbulence},}\ }\href@noop {} {\bibfield
  {journal} {\bibinfo  {journal} {Phys. Rev. Lett.}\ }\textbf {\bibinfo
  {volume} {115}} (\bibinfo {year} {2015})},\ \bibinfo {note}
  {124501}\BibitemShut {NoStop}%
\bibitem [{\citenamefont {Stone}\ \emph {et~al.}(2001)\citenamefont {Stone},
  \citenamefont {Brady},\ and\ \citenamefont {Lovalenti}}]{stone2016}%
  \BibitemOpen
  \bibfield  {author} {\bibinfo {author} {\bibfnamefont {H.~A.}\ \bibnamefont
  {Stone}}, \bibinfo {author} {\bibfnamefont {J.~F.}\ \bibnamefont {Brady}}, \
  and\ \bibinfo {author} {\bibfnamefont {P.~M.}\ \bibnamefont {Lovalenti}},\
  }\bibfield  {title} {\enquote {\bibinfo {title} {Inertial effects on the
  rheology of suspensions and on the motion of individual particles},}\
  }\href@noop {} {\  (\bibinfo {year} {2001})},\ \bibinfo {note} {submitted to
  J. Fluid Mech. (2016)}\BibitemShut {NoStop}%
\end{thebibliography}
\end{document}

\title{Supplemental material for: Angular dynamics of a small particle in turbulence}
\author{F. Candelier}
\affiliation{Aix-Marseille University - IUSTI (UMR CNRS 7343), 13 453 Marseille Cedex, France}
\author{J. Einarsson}
\affiliation{Department of Physics, Gothenburg University, SE-41296 Gothenburg, Sweden}
\author{B. Mehlig}
\affiliation{Department of Physics, Gothenburg University, SE-41296 Gothenburg, Sweden}
% \pacs{83.10.Pp,47.15.G-,47.55.Kf,47.10.-g}
% 83.10.Pp Particle dynamics (Fundamentals and theoretical)
% 47.15.G- Low-Reynolds-number (creeping) flows
% 47.55.Kf Particle-laden flows
% 47.10.-g General theory in fluid dynamics
\pacs{05.40.-a,47.55.Kf,47.27.eb,92.60.Mt}
% 05.40.-a Fluctuation phenomena, random processes, noise, and Brownian  motion
% 92.60.Mt Particles and aerosols
% 05.60.Cd Classical transport
% 45.50.Tn Collisions
% 47.27.-i Turbulent flows
% 05.40.Jc Brownian motion
% 47.55.Kf Particle-laden flows
% 47.27.eb Turbulence - Statistical theories and models
% 47.27.Gs Isotropic turbulence; homogeneous turbulence
\maketitle

All equations in this supplemental material are in dimensionless variables,
$ t = \tau_c  t^\prime\,, \ve r = a \ve r^\prime\,, \ve U = sa \ve U^\prime\,,$ and 
$\bbsigma = \mu s \bbsigma^\prime$.  For details see the Letter. For notational convenience we drop the primes in the following.

\section{Evaluation of Jeffery's torque}
Eq.~(10) in the Letter gives an expression for Jeffery's torque \cite{jeffery1922}:
\begin{align}
\label{eq:T0}
        \ve T^{(0)} &= -\ma K\cdot \ve \omega_{\rm s} + \ma H : \ma S_p^\infty\,,
\end{align}
written in terms of Brenner's resistance tensors $\ma K$ and $\ma H$ \cite{happel1983}. Here $\ve \omega_{\rm s}\equiv \ve \omega-\ve \Omega^\infty_p$
  is the angular slip velocity, $\ve \Omega^\infty_p = \tfrac{1}{2} \ve \nabla \wedge \ve U^\infty|_{\ve x_p}$, and $\ma S_p^\infty$ is the symmetric
part of the matrix $\ma A_p^\infty$ of the undisturbed fluid-velocity gradients evaluated at $\ve x_p$.  The anti-symmetric part is denoted
by $\ma O_p^\infty$. This matrix is related to $\ve \Omega^\infty_p$ by ${\ma O_p^\infty \cdot \ve r} = \ve \Omega_p^\infty\wedge \ve r$.
Expanding the resistance tensors in the nearly-spherical limit one finds \cite{kim1991}:
 \begin{equation}
\label{eq:jeffery}
 \begin{split}
 \ve T^{(0)}  =& -8\pi \big(1-\tfrac{9}{5} \epsilon +\tfrac{459}{350} \epsilon^2\big) \ve \omega_{\rm s} +8\pi \big(\tfrac{3}{5} \epsilon - \tfrac{177}{350} \epsilon^2\big) (\ve \omega_{\rm s} \cdot \ve n) \ve n 
 -8\pi  \big(\epsilon - \tfrac{13}{10} \epsilon^2\big)\big[ (\ma S_p^\infty \cdot \ve n)\wedge \ve n \big]\:.
 \end{split}
 \end{equation}
Here $\epsilon$ is a small parameter related to the particle eccentricity (see Ref.~\cite{candelier2015b}). For a sphere we have $K_{ij} = 8\pi \delta_{ij}$ and $\ma H=0$.

\section{Computation of disturbance torque using reciprocal theorem}
\label{sec:rt}
The hydrodynamic torque is given by
\begin{equation}
\label{eq:t0}
\ve T = \int_{S_p}\!\! \ve r \wedge  \bbsigma\cdot {\rm d}\ve s \,.
\end{equation}
The integral is over the particle surface $S_p$,  and
the vector ${\rm d}\ve s$ points in the direction of the outward surface normal. The stress tensor $\bbsigma$ is determined by the solution of the Navier-Stokes equations. In dimensionless variables it has elements $\sigma_{ij} = -\delta_{ij} p + 2 S_{ij}$ where
$p$ is the pressure and $S_{ij}$ are the elements of ${\ma S^\infty}$.
We decompose $\bbsigma$ as  $\bbsigma  = \bbsigma^\infty  + \bbsigma^{(1)}$,
where $\bbsigma^\infty$ is the undisturbed stress tensor and $\bbsigma^{(1)}$
is the contribution to the stress tensor from the disturbance flow. The hydrodynamic torque is split up in the same way:
\begin{equation}
\ve T^\infty = \int_{S_p}\!\! \ve r \wedge  \bbsigma^\infty\cdot {\rm d}\ve s \quad\mbox{and}\quad
\ve T^{(1)} = \int_{S_p}\!\! \ve r \wedge  \bbsigma^{(1)}\cdot {\rm d}\ve s \,.
\end{equation}
We use the reciprocal theorem \cite{lovalenti1993} to compute $\ve T^{(1)}$, 
following Refs.~\cite{subramanian2005,einarsson2015a,candelier2015b}. 
The reciprocal theorem relates integrals of two different Newtonian and incompressible flows:
\begin{equation}
\label{eq:rt0}
\int_{S_p}\! \ve w\cdot \tilde{\bbsigma}\cdot{\rm d}\ve s + \int_V\!{\rm d}v\, \ve w\cdot(\ve \nabla\cdot 
\tilde{\bbsigma})
= \int_{S_p}\! \tilde{\ve u}\cdot \bbsigma^{(1)}\cdot{\rm d}\ve s + \int_V\!{\rm d}v\, \tilde{\ve u}\cdot(\ve \nabla\cdot \bbsigma^{(1)})\,.
\end{equation}
Here $\ve w$ is the disturbance flow, it obeys Eq.~(9) in the Letter. The fields $\tilde{\ve u}$ and $\tilde\bbsigma$ are the fluid
velocity and the stress tensor of an \lq auxiliary\rq{} Stokes problem in the same geometry, namely the Stokes flow 
produced by particle rotating with
angular velocity $\tilde{\ve \omega}$  in a quiescent fluid:
\begin{subequations}
\begin{align}
\label{eq:aux}
& \boldsymbol{\nabla} \cdot \tilde{{\bbsigma}} = \ve 0\,, \\
& \tilde{\ve u} = \tilde{\ve \omega} \wedge \ve r\,\,\mbox{for} \,\, r \in S_p\quad\mbox{and}\quad \tilde {\ve u} \to 0 \,\,\mbox{as}\,\,r\to\infty\,.
\end{align}
\label{eq:testpb}
\end{subequations}
The hydrodynamic (auxiliary) torque on the particle is denoted by $\tilde {\ve T}$:
\begin{equation}
\tilde{\ve T} = -\ma K \cdot \tilde{\ve \omega}\,.
\end{equation}
The surface integrals in Eq.~(\ref{eq:rt0}) are over the particle surface, the vector ${\rm d}\ve s$ points in the direction of the outward surface normal. 
The volume integrals are  over the volume outside the particle. 

The surface integrals in Eq.~(\ref{eq:rt0}) are evaluated by substituting the boundary conditions
for the disturbance flow and the auxiliary flow on the particle surface.
The volume integral on the l.h.s. of  Eq.~(\ref{eq:rt0}) vanishes by virtue of Eq.~(\ref{eq:aux}). 
For the divergence of the disturbance stress tensor
in the volume integral on the r.h.s., we substitute Eq.~(9)
in the Letter. The vector $\ve f$ is defined by
\begin{equation}
\ve f \equiv \Sl \Big(\frac{\partial \ve w}{\partial t}\Big)_{\ve r} +
\ma A^\infty_p \cdot \ve w+
[(\ma A^\infty_p \cdot \ve{r}) \cdot \boldsymbol{\nabla} ] \ve{w} + (\ve{w}\cdot \ve \nabla) \ve{w}\:.
\end{equation}
We find:
\begin{equation}
\label{eq:rt01}
\int_{S_p}\!  (-\ma S_p^\infty\cdot \ve r+\ve \omega_{\rm s}\wedge \ve r )\cdot \tilde{\bbsigma}\cdot{\rm d}\ve s 
= \int_{S_p}\! (\tilde{\ve \omega}\wedge \ve r)\cdot \bbsigma^{(1)}\cdot{\rm d}\ve s 
+ \int_V\!{\rm d}v\, \tilde{\ve u}\cdot\ve f\,.
\end{equation}
The surface integrals can be expressed in terms of the torques $\ve T^{(1)}$ and $\tilde{\ve T}$:
\begin{equation}
\ve \omega_{\rm s}  \cdot \tilde{\ve T} 
- \int_{S_p} \left(\ma S^\infty_p \cdot \ve r \right) \cdot  \tilde{\bbsigma} \cdot {\rm d}\ve s  
= \tilde{\ve \omega} \cdot \ve T^{(1)}
+ \Reys\int_V {\rm d}v   \cdot \tilde{\ve u}\cdot\ve f
\label{eq:rt}
\end{equation}
This is an exact relation between the sought disturbance torque and the known auxiliary torque.
The disturbance velocity $\ve w$, however, is not known. To lowest order in $\Reys$ we can use the 
Stokes approximation $\ve w^{(0)}$ in evaluating the volume integral. 
The function $\ve w^{(0)}$ is the solution of the Stokes problem:
\begin{subequations}
\begin{align}
\boldsymbol{\nabla} \cdot {\bbsigma} &= \ve 0\,,\\
  \ve w^{(0)} &= \left(\ve \omega - \ve \Omega_p^{\infty}\right)  \wedge \ve r - \ma S_p^\infty \cdot \ve r  \,\, \mbox{for}\,\, r \in S_p\quad \mbox{and}\quad\ve w^{(0)} \to 0\,\,\mbox{as}\,\,r\to\infty\,.
\label{eq:pb}
\end{align}
\end{subequations}
The first two terms on the r.h.s. of (\ref{eq:rt}) evaluate to $\tilde{\ve \omega}\cdot \ve T^{(0)}$. Adding $\ve T^\infty$ [Eq.~(\ref{eq:tinfty})] we obtain Eq.~(10) in the Letter, with $\tilde {\ve u} = \ma M \cdot \tilde {\ve \omega}$. 

\section{Evaluation of torque and angular velocity}
\subsection{Sphere in a steady flow}
We first consider a simpler problem than that addressed in the Letter. What is the torque on a sphere
in the steady state where the sphere rotates with the steady angular velocity $\ve \omega = \ve \Omega_p^\infty + O(\Reys)$?
The undisturbed torque $\ve T^\infty$ follows from Eq.~(8) in the Letter
\begin{equation}
\label{eq:tinfty}
\ve T^\infty  = -\Reys \frac{8\pi}{15} \ma S_p^\infty \cdot \ve \Omega_p^\infty\,.
\end{equation}
The disturbance torque is evaluated using the reciprocal theorem (Section \ref{sec:rt}).
The solution of the auxiliary problem (\ref{eq:testpb}) for a spherical particle reads:
\begin{equation}
\tilde{u}_i  = \frac{1}{r^3} \varepsilon_{ijk} \tilde{\omega}_j r_k \equiv
M_{ij} \tilde \omega_j\,.
\label{eq:solutiontest}
\end{equation}
The last equality defines the elements of the matrix $\ma M$ used in Eq.~(8) in the Letter.
% Eq. (\ref{eq:solutiontest}) is the same as Eq.~(3.5b) in Howard's manuscript.
The corresponding torque $\tilde{\ve T}$  evaluates to 
\begin{equation}
\tilde{\ve T} = - 8\pi \tilde{\ve \omega} \,,
\end{equation}  
and the surface integral in Eq.~(\ref{eq:rt}) vanishes: 
\begin{equation}
\int_{S_p} \left(\ma S^\infty_p \cdot \ve r \right) \cdot  \tilde{\bbsigma} \cdot {\rm d}\ve s=0\:. 
\end{equation}
It remains to evaluate the volume integral in Eq.~(\ref{eq:rt}). The Stokes solution (the solution of (\ref{eq:pb}) 
for $\ve \omega = \ve \Omega_p^\infty$) reads:
\begin{equation}
w^{(0)}_i = -\frac{1}{r^5} (\ma S_p^\infty)_{ij} r_j - \frac{5}{2 } \frac{1}{r^5}\Big(1-\frac{1}{r^2}\Big) r_i r_j (\ma S_p^\infty)_{jk} r_k\,,
\end{equation}
The volume integral in Eq.~(\ref{eq:rt}) consists of four terms that we evaluate separately:
\begin{subequations}
\label{eq:IJK0}
\begin{align}    
I&= \int_V\!{\rm d}v\,   \Sl 
\Big(\frac{\partial \ve w^{(0)}}{\partial t}\Big)_{\ve r} \cdot \tilde{\ve u}\,,\\
J&= \int_V \!{\rm d}v \, ( \ma A^\infty_p  \cdot \ve w^{(0)} ) \cdot \tilde{\ve u} \,,\\
K&= \int_V \!{\rm d}v \, \big( [(\ma A^\infty_p \cdot \ve{r}) \cdot \boldsymbol{\nabla} ]
\ve w^{(0)}\big) \cdot  \tilde{\ve u} \,,\\
L&= \int_V \!{\rm d}v \, \big[(\ve w^{(0)}\cdot \ve \nabla)\ve w^{(0)}\big] \cdot  \tilde{\ve u} \,.
\end{align}
\end{subequations}
Since the problem is steady, the derivative $({\partial \ve w^{(0)}}/{\partial t})_{\ve r}$ vanishes, 
so that $I$ is zero:
\begin{equation}
\label{eq:I1}
I=0\:.
\end{equation}
The integral $J$ can be written in the form
\begin{equation}
\label{eq:J1}
J=- (\ma A_p^\infty)_{ij}  (\ma S_p^\infty)_{jk} \varepsilon_{i\ell m} \tilde{\omega}_\ell  \int_V {\rm d}v \frac{r_m r_k}{r^8} - \frac{5}{2} (\ma A_p^\infty)_{ij} (\ma S_p^\infty)_{kl}\varepsilon_{imn} \tilde{\omega}_m \int_V {\rm d}v \frac{r_j r_k r_\ell r_n}{r^8} \left(1-\frac{1}{r^2}\right) \:,
\end{equation}
The  two integrals in this equation  are evaluated using Maple:
\begin{equation}
\int_V {\rm d}v \frac{r_m r_k}{r^8} = \frac{4\pi}{9} \delta_{km} \quad \mbox{and} \quad \int_V {\rm d}v \frac{r_j r_k r_\ell r_n}{r^8} \left(1-\frac{1}{r^2}\right)  = \frac{8 \pi}{45} (\delta_{jk} \delta_{\ell n} + \delta_{j \ell}\delta_{k n} + \delta_{j n} \delta_{k \ell})\:.
\end{equation}
This yields
\begin{equation}
J = -\frac{4\pi}{3}  \varepsilon_{i\ell k} (\ma A_p^\infty)_{ij}  (\ma S_p^\infty)_{jk}  \tilde{\omega}_\ell\,.
\end{equation}
To evaluate this expression further we split $\ma A_p^\infty$ into its symmetric and anti-symmetric parts,
$\ma A_p^\infty=\ma S_p^\infty+\ma O_p^\infty$, and make use of the
relations
\begin{equation}
(\ma S_p^\infty)_{k \ell} \delta_{k \ell} = (\ma S_p^\infty)_{kk} = 0 \quad \mbox{(incompressibility)} \quad \mbox{and} \quad 
(\ma S_p^\infty)_{jn}= (\ma S^\infty_p)_{nj} \quad \mbox{(symmetry)}\,.
\end{equation}
We use 
\begin{align}
\varepsilon_{i \ell k} (\ma S_p^\infty) _{ij} (\ma S_p^\infty)_{j k} &= 0\,\quad\mbox{and}\quad \varepsilon_{i\ell k} (\ma O_p^\infty)_{ij}  = -\delta_{\ell j} \Omega_k + \delta_{kj} \Omega_\ell\,.
\end{align}
The second equation follows from
$(\ma O_p^\infty)_{ij} = -\varepsilon_{ijk} \Omega_k$ and 
$\varepsilon_{ijk} \varepsilon_{i\ell m}=\delta_{j\ell}\delta_{k m} - \delta_{j m}\delta_{k\ell}$.
We finally obtain:
\begin{equation}
\label{eq:J11}
J =  -\frac{4\pi}{3}  ( -\delta_{\ell j} \Omega_k + \delta_{kj} \Omega_\ell)
 (\ma S_p^\infty)_{jk}  \tilde{\omega}_\ell=
% \frac{4\pi}{3}  S_{\ell k} \Omega_k \tilde{\omega}_\ell  = 
 \frac{4\pi}{3}   \left(\ma S^\infty \cdot  \ve \Omega^\infty_p \right) \cdot  \tilde{\ve \omega}\:.
\end{equation}
The remaining integrals ($K$ and $L$) are more difficult to calculate.  
Using Maple we find:
%$$
%K=\int_v -A_{jk} r_k \left(S_{i \ell} \left(\frac{\delta_{j \ell}}{r^5} - 5 \frac{r_\ell r_j}{r^7} \right)
%+\frac{5}{2}  \left(-\frac{5}{r^7} + \frac{7}{r^9} \right) r_j r_i r_\ell r_m S_{m\ell}
%+\frac{5}{2} \left(\frac{1}{r^5} - \frac{1}{r^7} \right) ( \delta_{ij} r_\ell r_m +2S_{\ell j} r_\ell r_i  ) S_{m\ell}\right) \varepsilon_{inp} \tilde{\omega}_n r_p {\rm d}v
%$$
%and evaluates to 
\begin{equation}
\label{eq:KL1}
 K 
=  \frac{4\pi}{3}   (\ma S^\infty \cdot \ve  \Omega^\infty_p ) \cdot  \tilde{\ve \omega} \quad \mbox{and} \quad L =0\:.
\end{equation}
Taking these results together we find: 
\begin{equation}
\Reys\int_V {\rm d}v\,\tilde{\ve u}\cdot\ve f  = \Reys \left( \frac{8\pi}{3} \ma S^\infty_p \cdot \ve  \Omega^\infty_p\right)\cdot \tilde{ \ve \omega}\:. 
\end{equation}
This result is valid for an arbitrary choice of $\tilde {\ve \omega}$. It follows: 
\begin{equation}
\label{eq:tprime}
\ve T^{(1)} = -8\pi \left(\ve \omega - \ve \Omega_p^\infty\right) - \Reys  \frac{8\pi}{3} \ma S^\infty_p \cdot \ve  \Omega^\infty_p\:.
\end{equation}
Adding the torque due to
the undisturbed flow [Eq.~(\ref{eq:tinfty})] yields:
\begin{equation}
\ve T = \ve T^\infty+\ve T^{(1)}= -8\pi(\ve \omega-\ve \Omega_p^\infty) -\Reys 8\pi \tfrac{2}{5} \ma S^\infty_p \cdot \ve  \Omega^\infty_p
\end{equation}
This is Eq.~(11) in the Letter.
In the steady state the torque must vanish. This gives
\begin{equation}
\ve \omega = \ve \Omega_p^\infty -\tfrac{2}{5} \Reys \ma S^\infty_p \cdot \ve  \Omega^\infty_p\,.
\end{equation}
\subsection{Sphere in an unsteady flow}
As in the previous Section we assume that $\ve\omega = \ve \Omega_p^\infty+O(\Reys)$. But now $\ma S_p^\infty$ and $\ve \Omega_p^\infty$ depend on time. 
In evaluating the contribution from the volume integral perturbatively we must again set $\ve\omega = \ve\Omega_p^\infty$
in $\ve w^{(0)}$. We find:
\begin{equation}
I = 0\,.
\end{equation}
For non-spherical particles $I$ is not zero, see below. The remaining terms ($J, K$, and $L$)
remain unchanged. The torque to the disturbance flow [Eq. (12) in the Letter] now reads
\begin{equation}
\ve T^\infty =   \Reys \frac{8\pi}{15} 
\Big( \Sl \frac{\rD \ve \Omega^\infty}{\rD t} \Big|_{\ve x_p}
-  \ma S_p^\infty \cdot \ve \Omega_p^\infty\Big)\,.
\end{equation}
This gives
\begin{equation}
\ve T =  -8\pi(\ve \omega-\ve \Omega_p^\infty)+\Reys 8\pi\Big( \frac{1}{15}\Sl 
 \frac{\rD \ve \Omega^\infty}{\rD t} \Big|_{\ve x_p}
- \frac{2}{5}  \ma S_p^\infty \cdot \ve \Omega_p^\infty\Big)\,.
\end{equation}
In the unsteady case the torque does not vanish, in general. To determine
the angular velocity of the particle we must therefore solve Newton's equation,
Eq. (1) in the Letter. For the sphere ${\rm d}\ma I/{\rm d}t$ vanishes, so that we have
\begin{equation}
 \frac{\St}{15}  \frac{{\rm d}\ve \omega}{{\rm d}t}
=  -(\ve \omega-\ve \Omega_p^\infty) +\Reys \Big( \frac{\Sl}{15} 
 \frac{\rD \ve \Omega^\infty}{\rD t} \Big|_{\ve x_p}
- \frac{2}{5}  \ma S_p^\infty \cdot \ve \Omega_p^\infty\Big)\,.
\end{equation}
On the l.h.s. we can set ${\rm d}\ve \omega/{\rm d}t = 
{\rm d}\ve \Omega_p^\infty/{\rm d}t = {\rm D}\ve\Omega^\infty/{\rm D}t|_{\ve x_p}$. 
To order $O(\Reys)$ this gives
\begin{equation}
\label{eq:wfin}
\ve \omega = \ve \Omega_p^\infty + \frac{1}{15} (\Reys\Sl-\St)   \frac{\rD \ve \Omega^\infty}{\rD t} \Big|_{\ve x_p}
-\frac{2}{5}  \Reys  \ma S_p^\infty \cdot \ve \Omega_p^\infty\,.
\end{equation}
This is Eq.~(12) in the Letter. 

\subsection{Nearly spherical particle in unsteady flow} 
Now we consider a nearly spherical particle in an unsteady flow. This case is quite similar to the previous one, except that
now the particle is assumed to rotate with an angular velocity that is different from $\ve \Omega^\infty_p$. For a nearly spherical particle the angular slip velocity $\ve \omega_{\rm s}= \ve \omega-\ve \Omega_p^\infty$ is of order $\epsilon$ as we shall see. This follows from Jeffery's theory.

As the auxiliary problem we use the Stokes flow of a nearly spherical particle rotating with angular velocity $\tilde{\ve \omega}$ in a quiescent fluid.
The solutions of the auxiliary problem (and of the Stokes problem for $\ve w^{(0)}$) are found using perturbation theory in 
the small parameter $\epsilon$. Details of the method are given in the appendix of Ref.~\cite{candelier2015b}. 

We note that the problem to be solved is more difficult for a non-spherical particle than for a sphere. 
For a sphere the locus of the particle surface remains invariant under 
rotation.
This simplifies the problem substantially, because we can work with a basis that remains fixed in the laboratory frame.
For a non-spherical particle we need to consider how the locus of the particle surface changes. This may 
be  accomplished in a coordinate system that 
rotates with the particle. The time derivative of $\ve w^{(0)}$ in the volume integral is evaluated at fixed $\ve r$. This derivative must be
transformed to the rotating coordinate system:
\begin{equation}
\Sl \Big(\frac{\partial \ve w}{\partial t} \Big)_{\ve r} =  \Sl \Big(\frac{\partial  w_i'}{\partial t} \Big)_{\ve r'} \hat{\bf e}'_i(t) + \ve \omega \wedge \ve w - \left[\left(\ve \omega \wedge \ve r \right) \cdot \boldsymbol{\nabla}  \right] \ve w\,.
\label{eq:rot}
\end{equation}
The partial time derivative on the r.h.s. is evaluated at a fixed point $\ve r'$ relative to an observer rotating with the particle, $\hat{\bf e}'_i(t)$ are basis vectors rotating with the particle, and $w_i'$ are the components of $\ve w$ in that basis. We remark that the primes in Eq.~(\ref{eq:rot}) 
do not denote dimensionless variables.

The remaining three terms are also integrated in the rotating coordinate system (in this system the locus of $S_p$ does not change with time). Evaluating the volume integral in this way we find:
 \begin{equation}
 \begin{split}
 \label{eq:fpint}
  \int_V \!\!\rd v \,\ve f\cdot\tilde {\ve u}   = & 
   -8\pi 
   \Big(  -\tfrac{\Sl}{3} \frac{\rd \ve \omega_{\rm s}}{\rd t}
          \!-\!\tfrac{1}{3} \ma S_p^\infty \cdot \ve \Omega_p^\infty
          \!+\!\tfrac{1}{8}\: \ma S_p^\infty \cdot \ve \omega_{\rm s} \Big) \cdot\tilde{\ve \omega} \\
          &
  - 8 \pi \epsilon 
 \Big[\tfrac{628}{525}\ma S_p^\infty \cdot \ve \Omega_p^\infty         
 -\tfrac{32}{105} (\ve n\cdot \ma S_p^\infty\cdot \ve n) 
  \ve \Omega_p^\infty 
 \! + \tfrac{136}{525}(\ve n \cdot \ma S_p^\infty\cdot  \ve \Omega^\infty_p)\ve n\Big]\cdot\tilde{\ve \omega}
  \\
 &  -8 \pi \epsilon \Big\{
 \Big[\Big(-\tfrac{3}{5}\, \ma S_p^\infty \cdot {\ma O_p^\infty} -\tfrac{3\Sl}{5}   \frac{\rd\ma S_p^\infty}{\rd t} 
 + \tfrac{3}{35}\ma O_p^\infty \cdot \ma S_p^\infty - \tfrac{123}{280} \ma S_p^\infty \cdot  \ma S_p^\infty \Big)\cdot \ve      n\Big] \wedge \ve n \Big\}\cdot\tilde{\ve \omega}\,.
 \end{split}
 \end{equation}  
 In deriving this expression we neglected terms of the order of $\epsilon \ve \omega_{\rm s}$ and $\ve \omega_{\rm s}^2$.
 This is justified because we know that the slip angular velocity
 is of order $\epsilon$ in the creeping-flow limit, see below. 
Setting $\epsilon=0$
in Eq.~(\ref{eq:fpint}) we find again Eq.~(\ref{eq:IJK0}) derived in the previous Section using a different method (coordinate system fixed in the laboratory frame). We
note that $\ve \omega_{\rm s} =0$ to lowest order in $\Reys$ for $\epsilon=0$.

Returning to the non-spherical case we compute the angular velocity using the ansatz:
\begin{equation}
 \ve \omega = \ve \omega^{(0)}  + \epsilon \, \ve \omega^{(\epsilon)} +  \epsilon^2 \, \ve \omega^{(\epsilon^2)} +  
\St \, \ve \omega^{(\mbox{\scriptsize St})} 
+\Reys  \, \ve \omega^{(\mbox{\scriptsize Re}_{\rm s})}+ \epsilon\St \, \ve \omega^{(\epsilon \mbox{\scriptsize St})} 
+\epsilon \Reys  \, \ve \omega^{(\epsilon \mbox{\scriptsize Re}_{\rm s})}\,.
\end{equation}
To order $O(\epsilon)^2$ we obtain:
\begin{equation}
\label{eq:w_sphere}
\ve \omega^{(0)}  + \epsilon \, \ve \omega^{(\epsilon)} +  \epsilon^2 \, \ve \omega^{(\epsilon^2)}  = \ve \Omega_p^\infty  + \left(\epsilon +\tfrac{1}{2} \epsilon^2\right) \ve n \wedge \ma S^\infty_p\cdot \ve n\,.
\end{equation}
This is the second order of an expansion of Jeffery's equation
\cite{jeffery1922,candelier2015b} in the small parameter $\epsilon$.  
Now consider the corrections due to particle and fluid inertia.  For $\epsilon=0$ we find:
\begin{equation}
\label{eq:c4}
\Reys \ve \omega^{(\Reys)} + \St \ve \omega^{(\St)} =  \tfrac{1}{15} \left(\Reys\Sl - \St\right)  \tfrac{\rD \ve \Omega^\infty}{\rD t} \big|_{\ve x_p} - \tfrac{2}{5} \Reys \ma S_p^\infty \cdot \ve \Omega_p^\infty\:.
\end{equation}
This is again Eq.~(\ref{eq:wfin}). For non-spherical particles there are corrections to this expression. 
To simplify the form of the corrections we use the symmetries
of the problem summarised in Table~\ref{table:1}. 
\begin{narrowtext}
\begin{table}[t]
\caption{\label{table:1} Symmetries used in deriving Eq. (\ref{eq:corr}).}
\begin{tabular}{lr}
\hline\hline
  $\tr\ma S^\infty = 0$ & Incompressibility \\
  $\ma S^\infty = ({\ma S^\infty})^{\sf T}$ & Symmetry  \\
  $\ma O^\infty = -({\ma O^\infty})^{\sf T}$ & Antisymmetry  \\
%  $n_i \dot n_i = 0$ & Dynamics preserves magnitude \\
  $\omega(\ve  n) = \omega(-\ve n)$ \hspace*{2cm}& Inversion symmetry \\\hline\hline
\end{tabular}
\end{table}
\end{narrowtext}
We find that the $O(\epsilon\Reys)$-corrections to Eqs.~(\ref{eq:w_sphere}) and (\ref{eq:c4}) take the form
\begin{align} 
\label{eq:corr} \delta\ve \omega &\equiv \epsilon\Reys\,\ve \omega^{(\epsilon\Reys)} + \epsilon\St\,\ve \omega^{(\epsilon\St)}
=\beta_{1}  \tfrac{\rD \ve \Omega^\infty}{\rD t} \big|_{\ve x_p}
+ \beta_2 \ve n \cdot \tfrac{\rD \ve \Omega^\infty}{\rD t} \big|_{\ve x_p} \ve n
+ \beta_3  \big(\tfrac{\rd \ma S_p^\infty }{\rd t}\cdot \ve n\big)\wedge \ve n \\
&+\beta_{4} \ma S_p^\infty\cdot \ve \Omega_p^\infty
+ \beta_{5} [\ve n \cdot (\ma S_p^\infty\cdot \ve \Omega_p^\infty)] \ve n
+\beta_6  [(\ve n \cdot (\ma S_p^\infty\cdot\ve n)]\ve \Omega_p^\infty
+\beta_7 (\ve n \cdot \ve \Omega_p^\infty)(\ve \Omega_p^\infty\wedge \ve n) \nonumber\\
&+\beta_8 (\ma S_p^\infty \cdot \ma S_p^\infty \cdot \ve n)\wedge \ve n
+ \beta_9 [\ma S_p^\infty \cdot (\ve \Omega_p^\infty\wedge \ve n)]\wedge \ve n + \beta_{10}  (\ve n \cdot \ve \Omega_p^\infty) \:\ma S_p^\infty \cdot \ve n \,.\nonumber
\end{align}
For $\Sl=1$ the coefficients read:
\begin{align}
\beta_{1} &= -\tfrac{3\epsilon}{15}(\Reys-\St)\,,\,\,\beta_{2} = -\tfrac{\epsilon}{15}(\Reys-\St)\,,\,\, 
\beta_3 = -\tfrac{\epsilon}{3} (\Reys - \tfrac{1}{5}\St) \,,\,\, \beta_{4}  = \tfrac{733\epsilon}{525}  \Reys\,,\\
\beta_{5} &=-\tfrac{\epsilon}{15} (\tfrac{4}{35}\Reys+ \St)\,, \beta_6 = - \tfrac{\epsilon}{15}(\tfrac{30}{35} \Reys-\St)\,,\,\,
\beta_7 = -\tfrac{\epsilon}{15}(\Reys - \St)\,,\,\, \beta_8 = -\tfrac{8\epsilon}{21} \Reys\,,\nonumber\\
\beta_9 &=-\tfrac{\epsilon}{15}(4\Reys - \St)\,,\,\, \beta_{10}= \tfrac{3\epsilon}{35} \Reys\,.\nonumber
\end{align}
We see that the inertial correction to the angular velocity of a non-spherical particle 
depends intricately on the alignment of % the particle symmetry vector $\ve n$, 
the vorticity $\ve \Omega_p^\infty(t)$ %, 
and the local eigensystem
$\hat{\bf s}_j(t)$ of $\ma S_p^\infty(t)$. In perturbation theory, to lowest order
in $\Reys$,  all time derivatives are evaluated as Lagrangian time 
derivatives along fluid trajectories.

For a neutrally buoyant particle in incompressible isotropic homogeneous turbulence we can average the correction.
Since $\delta\ve \omega$ contains a factor $\epsilon$ we obtain the correct result to order $O(\epsilon\Reys)$  by averaging $\ve n$ uniformly over the unit sphere,
${\langle n_i  n_j\rangle} = \delta_{ij}/3$. Using that $\langle \tr\ma S\cdot \ma O\rangle = \langle \tr\ma O\cdot \ma O\cdot \ma O\rangle =0$ and $\langle\tr\ma S\ma \cdot S\rangle = \langle\tr\ma O\ma \cdot O\rangle$ we find
\begin{equation}
\langle \ve \Omega_p^\infty\cdot \delta\ve \omega\rangle = \tfrac{4}{3} \epsilon\Reys  \langle \ve\Omega_p^\infty\cdot(\ma S\cdot\ve \Omega_p^\infty)\rangle
\,.
\end{equation}
This result is discussed in the Letter.